\documentclass[useAMS,usenatbib]{mnras}
\usepackage{aasmacros,placeins,float,times}
\usepackage{amsmath}
\usepackage{graphicx,url,pifont}
\usepackage{amssymb,color}
\def\gsim{\mathrel{\hbox{\rlap{\hbox{\lower4pt\hbox{$\sim$}}}\hbox{$>$}}}}

\newcommand{\Apostle}{\textsc{apostle}}
\newcommand{\Auriga}{\textsc{auriga}}
\newcommand{\Eagle}{\textsc{eagle}}
\newcommand{\Illustris}{\textsc{illustris}}
\newcommand{\IllustrisTNG}{\textsc{illustrisTNG}}

\newcommand{\PGadget}{\textsc{p-gadget-3}}
\newcommand{\Anarchy}{\textsc{anarchy}}
\newcommand{\Subfind}{\textsc{subfind}}
\newcommand{\Arepo}{\textsc{arepo}}
\newcommand{\Fire}{\textsc{fire}}
\newcommand{\ApostleHydro}{\textsc{apostle-hydro}}
\newcommand{\ApostleDMO}{\textsc{apostle-dmo}}
\newcommand{\AurigaMHD}{\textsc{auriga-mhd}}
\newcommand{\AurigaDMO}{\textsc{auriga-dmo}}

\newcommand{\bq}{\begin{eqnarray}}
\newcommand{\eq}{\end{eqnarray}}

\title[SFHs, cores and cusps]{No cores in dark matter-dominated dwarf
  galaxies with bursty star formation histories}

\author[S. Bose et al.]{Sownak Bose$^{1}$\thanks{Email:
    sownak.bose@cfa.harvard.edu}, Carlos S. Frenk$^{2}$, Adrian
  Jenkins$^{2}$, Azadeh Fattahi$^{2}$, \newauthor Facundo
  A. G{\'o}mez$^{3,4}$, Robert J. J. Grand$^{5,6}$, Federico
  Marinacci$^{1}$, \newauthor Julio F. Navarro$^{7}$, Kyle
  A. Oman$^{8,7}$, R{\"u}diger Pakmor$^{9}$, Joop Schaye$^{10}$,
  \newauthor Christine M. Simpson$^{11,12,5}$, and Volker
  Springel$^{9,5,6}$ \\ \\ $^{1}$ Harvard-Smithsonian Center for
  Astrophysics, 60 Garden St., Cambridge, MA 02138, USA \\ $^{2}$
  Institute for Computational Cosmology, Durham University, South
  Road, Durham, DH1 3LE, UK \\ $^{3}$ Instituto de Investigaci{\'o}n
  Multidisciplinar en Ciencia y Tecnolog{\'i}a, Universidad de La
  Serena, Ra{\'u}l Bitr{\'a}n 1305, La Serena, Chile \\ $^{4}$
  Departamento de F{\'i}sica y Astronom{\'i}a, Universidad de
  LaSerena, Av. Juan Cisternas 1200 N, La Serena, Chile \\ $^{5}$
  Heidelberger Institut f{\"u}r Theoretische Studien,
  Schlo{\ss}-Wolfsbrunnenweg 35, 69118 Heidelberg, Germany \\ $^{6}$
  Zentrum f{\"u}r Astronomie der Universit{\"a}t Heidelberg,
  Astronomisches Recheninstitut, M{\"o}nchhofstr. 12-14, 69120
  Heidelberg, Germany \\ $^{7}$ Department of Physics and Astronomy,
  University of Victoria, PO Box 3055 STN CSC, Victoria, BC, V8W 3P6,
  Canada \\ $^{8}$ Kapteyn Astronomical Institute, University of
  Groningen, Postbus 800, NL-9700 AV Groningen, The Netherlands
  \\ $^{9}$ Max-Planck-Institut f{\"u}r Astrophysik,
  Karl-Schwarzschild-Str. 1, D-85748, Garching, Germany \\ $^{10}$
  Leiden Observatory, Leiden University, PO Box 9513, 2300 RA Leiden,
  The Netherlands \\ $^{11}$ Enrico Fermi Institute, The University of
  Chicago, Chicago, IL 60637, USA \\ $^{12}$ Department of Astronomy
  \& Astrophysics, University of Chicago, Chicago, IL 60637, USA}
\begin{document}

\pagerange{\pageref{firstpage}--\pageref{lastpage}} \pubyear{2018}

\maketitle

\label{firstpage}

\begin{abstract}
Measurements of the rotation curves of dwarf galaxies are often
interpreted as requiring a constant density core at the centre, at
odds with the ``cuspy'' inner profiles predicted by $N$-body
simulations of cold dark matter (CDM) haloes. It has been suggested
that this conflict could be resolved by fluctuations in the inner
gravitational potential caused by the periodic removal of gas
following bursts of star formation. Earlier work has suggested that
core formation requires a bursty and extended star formation history
(SFH). Here we investigate the structure of CDM haloes of dwarf
galaxies ($M_{{\rm DM}} \sim 10^9-5\times10^{10}\,{\rm M}_\odot$)
formed in the \Apostle{} (`A Project of Simulating the Local
Environment') and \Auriga{} cosmological hydrodynamic simulations. Our
simulations have comparable or better resolution than others that make
cores ($M_{{\rm gas}} \sim 10^4\,{\rm M}_\odot$, gravitational
softening $\sim 150$ pc). Yet, we do not find evidence of core
formation at {\it any} mass or any correlation between the inner slope
of the DM density profile and temporal variations in the
SFH. \Apostle{} and \Auriga{} dwarfs display a similar diversity in
their cumulative SFHs to available data for Local Group dwarfs. Dwarfs
in both simulations are DM-dominated on all resolved scales at all
times, likely limiting the ability of gas outflows to alter
significantly the central density profiles of their haloes. We
conclude that recurrent bursts of star formation are not sufficient to
cause the formation of cores, and that other conditions must also be
met for baryons to be able to modify the central DM cusp.
\end{abstract}

\begin{keywords}
   cosmology: dark matter -- galaxies: dwarf -- galaxies: haloes -- galaxies: Local Group -- galaxies: star formation
 \end{keywords}

\section{Introduction}
\label{sect:Intro}

The existence of dark matter (DM) in the form of cold, collisionless
particles is the bedrock of the currently favoured model of cosmology,
$\Lambda$CDM. In this model, the accelerated expansion of the Universe
on large scales is dominated by vacuum energy in the form of a
cosmological constant, $\Lambda$, while structure formation on small
scales proceeds hierarchically through the gravitational collapse of
cold dark matter (CDM) particles into DM ``haloes''.  The theory of
galaxy formation, which has matured over the last four decades, has
painted a picture where baryons are able to cool and condense into
these DM haloes, eventually forming the stars that make up a galaxy
\citep{WhiteFrenk1991}. The death of massive stars in the form of
supernovae releases energy back into the surrounding gas, reheating it
to suppress further star formation, before radiative cooling of this
heated gas is able to kick-start star formation once again
\citep[e.g.][]{Larson1974,Dekel1986,Katz1996,SomervillePrimack1999,Cole2000}.

A feature of the CDM model that has enhanced its prominence is that it
is highly predictive. Many of its predictions, particularly in the
non-linear regime of structure formation, have come from an intensive
programme of $N$-body simulations over the past three decades
\cite[see][for a recent review]{FrenkWhite2012}. A fundamental
prediction from collisionless $N$-body simulations is that DM haloes
develop density profiles with steeply rising slopes in the inner part
of the halo, described by the Navarro-Frenk-White (NFW) density
profile \citep{NFW1996,NFW1997}. This profile rises as $\rho \propto
r^{-1}$ in the centre, resulting in a central ``cusp'', as $\rho
\propto r^{-3}$ in the outer parts, and as $\rho \propto r^{-2}$ in
between. The NFW profile is universal \citep[i.e. independent of halo
  mass, but see e.g.][for claimed deviations at much smaller mass
  scales]{Anderhalden2013,Ishiyama2014,Angulo2017}.

In conjunction with simulations, our understanding of the Universe
around us has also been augmented by the exquisite observational data
now available, especially for galaxies in the Local
Group. DM-dominated dwarf galaxies, in particular, are ideal for
investigating the interplay between the gravitational collapse of DM
and the physics of galaxy formation. These investigations, however,
have not been without controversy. It has been claimed that the DM
density profiles of dwarf galaxies, inferred from their H\,{\sc i}
rotation curves or stellar kinematics, reveal the presence of a near
constant density inner ``core'', in stark contrast with the prediction
of the NFW model
\citep{Moore1994,Flores1994,Burkert1995,deBlok2001,deNaray2011,Hague2013,Oh2015}.
This mismatch between theory and observation, the so-called {\it
  core-cusp problem}, is often cited as one of the greatest challenges
faced by the CDM paradigm.

In reply, theorists have proposed mechanisms to induce cores in
originally cuspy profiles. The main idea goes back to the work of
\cite{NEF1996} who showed that a core can be produced by the sudden
removal of gas (by energy injected from supernovae) from the centre of
a cuspy halo in which gas had previously cooled gradually until
dominating the gravitational potential. To illustrate this mechanism
they assumed an initial analytic mass distribution corresponding to a
cuspy density profile\footnote{\cite{NEF1996} used the cuspy Hernquist
  profile \citep{Hernquist1990} to represent the DM density
  distribution.}  which was perturbed by the potential of a gradually
growing baryonic disk. To mimic the effect of an energetic outflow,
the disk potential was removed abruptly; the DM responds to this
change by settling into a new equilibrium configuration with a central
core whose size depends on the strength of the perturbation.

The idea that energetic outflows may generate cores was further
developed by \cite{Read2005} and \cite{Mashchenko2006,Mashchenko2008}
who argued that a {\it series} of localised, moderately violent
outbursts, is a more efficient way of generating a core than the
single, explosive outburst mechanism of \cite{NEF1996}. The process
was first seen in cosmological hydrodynamic simulations by
\cite{Governato2010} and \cite{Parry2012}, and the physics behind core
creation through repeated outbursts was later detailed by
\cite{Pontzen2012}. Their proposed model describes oscillations in the
gas potential generated by repeated bursts that eventually transfer
energy to the DM, expanding the orbits of particles near the halo
centre, transforming a cusp into a core. \cite{Governato2010} also
found that the efficacy of this mechanism depends on the threshold
density for star formation, $n_{{\rm sf}}$, assumed in the
simulation. A low threshold ($n_{{\rm sf}} = 0.1$ cm$^{-3}$) preserves
a cusp, while a high threshold ($n_{{\rm sf}} = 100$ cm$^{-3}$) leads
to a core. More recently, \cite{Zant2016} have proposed a theoretical
framework for understanding the mechanisms for core formation in terms
of statistical properties of fluctuations in the gaseous component of
the halo.

Several hydrodynamical simulations have reported a connection between
the formation of cores and the star forming efficiency of dwarf
galaxies. For example, \cite{DiCintio2014,Tollet2016}
and~\cite{Maccio2017} find a strong dependence of the inner slope of
the DM density profile on the final stellar-to-halo mass ratio,
$M_\star/M_{{\rm h}}$. Galaxies in which star formation is inefficient
($M_\star/M_{{\rm h}} \lesssim 10^{-4}$), do not form cores;
conversely, highly star forming galaxies ($M_\star/M_{{\rm h}} \gtrsim
10^{-2}$) develop even cuspier profiles than their DM-only
counterparts due to adiabatic contraction
\citep[e.g.][]{Duffy2010,Schaller2015}. These limits bracket a
``sweet-spot'' for core creation at $M_\star/M_{{\rm h}} \sim
10^{-2}$. An interesting result of these works is that the {\it
  qualitative} relationship between the inner slope of the profile and
$M_\star/M_{{\rm h}}$ is seemingly independent of the specific
feedback implementation in the simulations.

Using the \Fire{} simulations \citep{Hopkins2014,Hopkins2017},
\cite{Onorbe2015} and \cite{Chan2015} found that while all their
simulated dwarfs exhibited extremely bursty SFRs (i.e. showing $\sim$
order of magnitude fluctuations in the SFR over a dynamical time), the
ones that preferentially formed cores were those with a substantial
amount of late-time star formation \citep[a similar observation has
  also been made more recently by][]{Read2018}. This stems primarily
from the fact that haloes that form cores during early bursts of star
formation are subject to many subsequent events of mass accumulation
through mergers and smooth accretion \citep[during what is known as
  the `rapid accretion phase'; see e.g.][]{Wechsler2002}. The result
of this is that `transient' cores are formed, which eventually
reassemble into cusps through these accretion events
\citep[e.g.][]{Laporte2015}. The requirements for core formation were
refined further by \cite{Fitts2017}, who corroborated the limit of
$\sim 10^6 {\rm M}_\odot$ as the `threshold' stellar mass needed to
form cores in dwarf galaxy haloes as previously reported by
e.g. \cite{Madau2014}. In other words, these authors find that dwarf
galaxies that exhibit the highest {\it star formation efficiency} have
the greatest propensity to form cores.

Other authors have proposed more exotic alternatives to CDM in which
the dynamics of the particles lead naturally to core formation on the
mass scales of interest. The most popular amongst these is warm dark
matter \citep[WDM,][]{Bond1983,Colin2000,Bode2001}. The free-streaming
of WDM particles suppresses density fluctuations below a
characteristic mass scale imposing constraints on the available
phase-space for the DM particles that result in the formation of a
core. However, \cite{VillaescusaNavarro2011} and \cite{Shao2013} have
shown that for WDM models that are observationally viable, the cores
are too small to be astronomically interesting, a result seen in
recent cosmological simulations where the overall NFW shape is
preserved on the scales of interest \citep[see,
  e.g.][]{Lovell2014,Bose2016,Bozek2016}. A more promising alternative
are self-interacting DM models, where multiple scattering events
between DM particles can result in the formation of constant density
cores by removing particles from the centres of haloes
\citep[e.g.][]{Vogelsberger2012,Zavala2013,Rocha2013,Robertson2018}.

Our objective in this paper is to examine the link, if any, between
the shape of the DM density profiles of dwarf galaxy haloes and their
SFHs in cosmological, hydrodynamical simulations of Milky Way and
Local Group-like environments. We investigate dwarf galaxies extracted
from the \Apostle{} \citep{Fattahi2016,Sawala2016} and \Auriga{}
\citep{Grand2017} projects. An important feature of the galaxy
formation models implemented in these simulations is that very similar
subgrid prescriptions have been shown to reproduce a wide variety of
properties of the galaxy population as a whole, such as the stellar
mass function of galaxies, the bimodality of their colour
distributions, etc.
\citep[e.g.][]{Schaye2015,Trayford2017,Pillepich2018b,Nelson2018}. This
point, and more specific details of these simulations, are elaborated
on in Section~\ref{sect:Sims}.

This paper is organised as follows. In Section~\ref{sect:Sims}, we
introduce the simulations used in this work and outline the criteria
to select an appropriate sample of dwarf galaxies
(Section~\ref{sect:Definitions}). Section~\ref{sect:Results} presents
our main results: the DM density profiles of dwarf galaxy haloes and
the evolution of these profiles in time (Section~\ref{sect:DMprof});
the bursty star formation rates of our simulated dwarfs and the SFHs
of our sample compared with observational data
(Section~\ref{sect:SFHistories}). In Section~\ref{sect:coreFormation},
we discuss possible reasons why our simulations do not form cores at
any mass. Finally, our conclusions are summarised in
Section~\ref{sect:Conclusions}.

\section{Simulations}
\label{sect:Sims}

In this section, we provide brief descriptions of \Apostle{} and
\Auriga{}, which are the sets of hydrodynamical simulations analysed
in this paper.

\subsection{The \Apostle{} simulations}
\label{sect:APOSTLE}

The \Apostle{} (`A Project Of Simulating The Local Environment')
simulation suite consists of a set of zoom-in hydrodynamical
simulations representing analogues of the Local Group and its
environment \citep{Fattahi2016,Sawala2016}. Pairs of haloes with total
mass, separation, and relative radial and tangential velocities
consistent with the Milky Way-M31 pair were selected from a periodic,
cosmologically representative dark matter only (DMO) simulation with a
comoving box size of 100 Mpc. The selected regions were then
re-simulated at higher resolution. The cosmological parameters used in
both the parent volume and each of the \Apostle{} re-simulations are
consistent with WMAP-7 \citep{Komatsu2011}: $\Omega_m = 0.272$,
$\Omega_b = 0.0455$, $\Omega_\Lambda = 0.728$ and $h = 0.704$, where
$h$ is related to the present day Hubble constant, $H_0$, by $h =
H_0/100{\rm kms}^{-1}{\rm Mpc}^{-1}$. The spectral index of the
primordial power spectrum, $n_s = 0.967$; the linear power spectrum is
normalised at $z=0$ using $\sigma_8 = 0.81$.

In total, 12 regions were selected for re-simulation as part of the
\Apostle{} simulation suite. While all 12 volumes were re-simulated at
`low' and `medium' resolution (L3 and L2), six \Apostle{} volumes have
also been run at `high' resolution (L1), three of which are used in
the present analysis (which we will label `Ap-V1', `Ap-V4' and `Ap-V6'
in the rest of this paper). In the \Apostle{} L1 simulations, a single
dark matter particle has a mass of $m_{{\rm DM}} \sim 4 \times 10^4
{\rm M}_\odot$, a single gas particle initially has an average mass of
$m_{{\rm gas}} \sim 7.4 \times 10^3 {\rm M}_\odot$, while the
gravitational softening at $z=0$ is set to $\epsilon = 134$
pc\footnote{These are representative values; in detail, they vary
  slightly from volume to volume.}. The results presented in this
paper use the \Apostle{} L1 simulations only; however, we have checked
explicitly that the results are converged at L2 and L3.

The \Apostle{} project was performed using the \Eagle{} simulation
code \citep{Schaye2015,Crain2015}, a modified version of the massively
parallel smoothed particle hydrodynamics (SPH) code, \PGadget{}
\citep{Springel2005b,Springel2008}. The \Eagle{} code contains several
updated subgrid physics models for the cooling and heating of gas
\citep{Wiersma2009a}; star formation and reionisation
\citep{Schaye2004,Schaye2008}; stellar mass loss and enrichment
\citep{Wiersma2009b}, as well as the feedback from stars and AGN
\citep{Booth2009,DallaVecchia2012}. A comprehensive discussion of the
subgrid prescriptions and the effect of varying their parameters can
be found in \cite{Schaye2015} and \cite{Crain2015}. SPH quantities and
hydrodynamic forces are computed using the \Anarchy{} SPH scheme (see
\citealt{Schaller2015b} for details), itself based on the
pressure-entropy SPH formulation described in \cite{Hopkins2013}. For
the conversion of gas into stars, a density threshold $n_{{\rm sf}} =
0.1 \left( Z/0.002 \right)^{-0.64}$ cm$^{-3}$ is adopted in
\Apostle{}, where $Z$ is the gas metallicity. Furthermore, because the
simulation is unable to adequately resolve or model the cold phase of
the interstellar medium (ISM), a temperature floor of $\sim 10^4$ K is
adopted, imposing an effective equation of state on the unresolved
ISM. Finally, we note that the parameters for the subgrid
implementation in the \Apostle{} project correspond to the \Eagle{}
{\sc reference} model.

\begin{table}
\centering
\vspace{10pt}
\begin{tabular}{cccc}
\hline \hline \\
{\bf Simulation} & {\bf Volume} &  ${\mathbf N_{{\rm dwarf}}}$ & ${\mathbf N_{{\rm dwarf}}}$ \\
           &        & [all; $z=0$]      & [luminous; $z=0$] \\
\hline \\
\Apostle{}: {\sc hydro} \& {\sc dmo} & (1) & (2) & (3) \\
\hline
& Ap-V1     & 146 & 62  \\
& Ap-V4     & 240 & 83  \\
& Ap-V6     & 240 & 89  \\
\hline
\Auriga{}: {\sc mhd} \& {\sc dmo} & & \\
\hline
& Au-6  & 17  & 14  \\
& Au-16 & 35  & 31  \\
& Au-21 & 30  & 29  \\
& Au-23 & 19  & 19  \\
& Au-24 & 51  & 46  \\
& Au-27 & 26  & 24  \\
\\ \hline \hline
\end{tabular}
\caption{Number of isolated dwarf galaxies (see definition in
  Section~\ref{sect:Definitions}) identified in the \Apostle{} and
  \Auriga{} simulations. Column (2) lists {\it all} dwarf galaxy
  haloes in the appropriate mass range; column (3) lists the number of
  them that are {\it luminous}, i.e. those that have formed at least
  one star particle. The larger simulation volume in \Apostle{}
  compared to \Auriga{} results in the presence of many more candidate
  dwarf haloes.}
\label{tab:dwarfnums}
\end{table}

\subsection{The \Auriga{} simulations}
\label{sect:Auriga}

The \Auriga{} project \citep{Grand2017} focuses specifically on
re-simulations of Milky Way mass haloes, rather than the Local Group
environment. Re-simulation candidates were chosen from the same 100
Mpc periodic box as the \Eagle{} project. To ensure a relatively
isolated sample of Milky Way-like systems, candidate haloes were
required to have a present-day mass \footnote{Here, the mass,
  $M_{200}$, is defined as the mass contained within the radius,
  $r_{200}$, which encompasses a mean matter density equal to 200
  times the critical density of the Universe at a given redshift.}
$10^{12}~<~ M_{200}/{\rm M}_\odot~<~2~\times~10^{12}$. The centre of a
target halo is also required to be located outside $9\times r_{200}$
of any other halo that has a mass greater than 3 per cent of the
target halo mass. The parent volume and subsequent re-simulations
assume cosmological parameters derived by {\it Planck}
\citep{Planck2014}: $\Omega_m = 0.307$, $\Omega_b = 0.04825$,
$\Omega_\Lambda = 0.693$, $h = 0.6777$, $n_s = 0.9611$ and $\sigma_8 =
0.8288$. The cosmological parameters and input power spectrum are
exactly the same as those used in the \Eagle{} project.

In total, 30 candidate haloes were selected for re-simulation: while
all 30 have LR and MR versions, six of them have been re-simulated at
high-resolution \citep[HR, corresponding to `Level 3' in the
  nomenclature of][]{Grand2017}. In this paper, these six haloes will
be labelled as `Au-6', `Au-16', `Au-21', `Au-23', `Au-24' and
`Au-27'. The HR \Auriga{} simulations are specified by $m_{{\rm DM}} =
4 \times 10^4 {\rm M}_\odot$, $m_{{\rm gas}} \sim 6 \times 10^3 {\rm
  M}_\odot$ and $\epsilon = 184$ pc. Nominally, the numerical
resolution of both \Apostle{} and \Auriga{} is comparable to or better
than that of other works in the literature, which do report cores.

A significant difference between \Apostle{} and \Auriga{} is that
while the former uses the SPH approach to solve the hydrodynamics,
\Auriga{} makes use of the magnetohydrodynamics (MHD) code, \Arepo{}
\citep{Springel2010}, which implements a moving, unstructured Voronoi
mesh to solve the MHD equations \citep{Pakmor2014}. In this sense,
$m_{{\rm gas}}$ in \Auriga{} refers to mass associated with a
particular gas cell in the Voronoi mesh, rather than to the mass of an
SPH particle. The moving mesh in \Auriga{} is adaptive, resolving
regions of high density with many more cells of a smaller size than in
low density environments.

In addition to the different approach to solving the hydrodynamics,
the subgrid implementation in \Auriga{} is also somewhat different,
deriving primarily from the treatment of gas cooling and heating, star
formation, metal enrichment, stellar and AGN feedback laid out in
\cite{Vogelsberger2013}, \cite{Marinacci2014} and
\cite{Pillepich2018a}\footnote{Note that while stellar winds are
  treated as in the \IllustrisTNG{} model \citep{Pillepich2018a},
  \Auriga{} uses the AGN prescription from the original \Illustris{}
  model. We do not expect AGNs to play a significant role in the
  present analysis.}. The density threshold for star formation
$n_{{\rm sf}}$ = 0.13 cm$^{-3}$ in \Auriga{}, although, unlike in
\Apostle{}, there is no explicit dependence of this threshold on the
metallicity of the star forming gas. As in \Apostle{}, a temperature
floor of $\sim 10^4$ K is also adopted. The \Auriga{} model also
includes a simple prescription for the self-shielding of dense gas
($>10^{-3}$ cm$^{-3}$) from background ultraviolet radiation;
self-shielding is not modelled in \Apostle{}.

There are also differences in the manner in which supernova feedback
is implemented in the respective models. \Apostle{} follows the scheme
outlined in \cite{DallaVecchia2012}, in which energy from supernovae
is dumped stochastically in a thermal component only, resulting in a
constant temperature increase of gas particles receiving this energy
by an amount $\Delta T = 10^{7.5}$ K. The resulting energy injected
per stellar mass formed depends on local properties of the gas
(i.e. its density and metallicity). On the other hand, \Auriga{} uses
the method of \cite{Marinacci2014} to deposit feedback energy as
kinetic and thermal components in equal parts. This feedback is
modelled by converting gas cells in wind particles, where the wind
velocity is set to 3.64$\sigma_{{\rm DM}}^{{\rm 1D}}$; here
$\sigma_{{\rm DM}}^{{\rm 1D}}$ is the local 1D DM velocity dispersion
\citep[c.f.][]{Okamoto2010}.

Finally, we note that every volume re-simulated as part of the
\Apostle{} and \Auriga{} projects have DMO counterparts simulated from
the same set of initial conditions. This is particularly important as
our goal is to study the effect of galaxy formation physics on the
inner structure of dark matter haloes compared to collisionless
simulations.

\begin{figure*}\centering \includegraphics[trim={0mm 0mm 0mm 0mm},width=\textwidth]{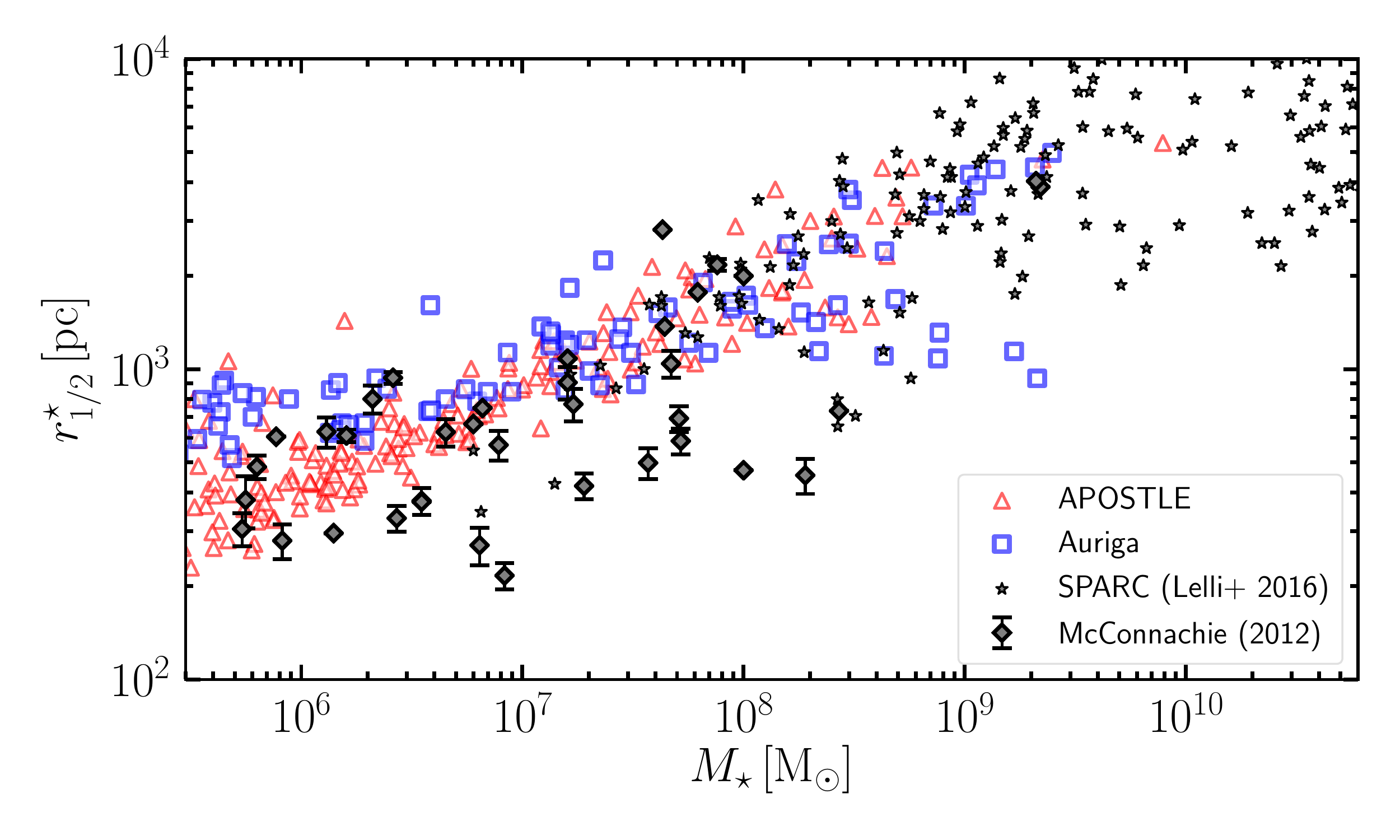}
  \caption{Galaxy stellar half-mass radius, $r_{1/2}^{\star}$, versus
    stellar mass, $M_{\star}$, for isolated galaxies identified in the
    three high-resolution \Apostle{} volumes, and the six
    high-resolution \Auriga{} volumes. The stellar mass of each galaxy
    is defined as the total mass in stars bound to the halo as
    determined by \Subfind{}. The grey diamonds with error bars show
    the published data compiled for the isolated Local Group dwarfs by
    \citet{Mcconnachie2012}, while the stars represent galaxies from
    the SPARC sample compiled by \citet{Lelli2016}.}
\label{fig:galaxySizes}
\end{figure*}

\subsection{Definitions and sample selection}
\label{sect:Definitions}

A post-processing step common to both \Apostle{} and \Auriga{} is the
identification of haloes and subhaloes. First, haloes are identified
using the `friends-of-friends' (FOF) algorithm, in which dark matter
particles separated by at most 0.2 times the mean inter-particle
separation are linked together to form groups
\citep{Davis1985}. Within each group, sets of gravitationally bound
substructures are identified using the \Subfind{} algorithm
\citep{Springel2001}. This splits a FOF halo into a `main' halo and
its associated subhaloes: one can think of this as the distinction
between the hosts of `central' and a `satellite' galaxies. In what
follows, we will be concerned with the `main' halo of FOF groups
only. We determine the centres of haloes using the {\it shrinking
  sphere} method \citep[e.g.][]{Power2003}, which identifies the
density maximum of a self-bound structure by recursively computing the
centre of mass of all DM particles within a shrinking sphere, until a
convergence criterion is met. In each iteration, the radius of the
sphere is reduced by $5$ per cent, and stopped when only 1000
particles or $1$ per cent of the particles of the initial sphere
(whichever is smaller) are left. In the vast majority of cases, the
shrinking sphere centre coincides with the location of the particle
with the minimum value of the gravitational potential identified by
\Subfind{}.

In what follows, we will be concerned primarily with the haloes of
{\it isolated} dwarf galaxies. Isolated (or `field') haloes are
objects found at a distance greater than 300 kpc away from the main
galaxy (i.e. the Milky Way analogue). In the case of \Apostle{}, we
require an isolated halo to be more than 300 kpc away from both the
Milky Way {\it and} M31 analogues. As these criteria are enforced at
$z=0$, our selection will inevitably include a small fraction ($\sim
20\%$) of ``backsplash'' galaxies: those that were once part of a
larger host, but are not any longer.  A dwarf galaxy is defined as
being in the mass range $10^9 < M_{{\rm DM}}/{\rm M}_\odot < 5 \times
10^{10}$, where $M_{{\rm DM}}$ is the bound DM mass associated with
the isolated galaxy as identified by \Subfind{}. The properties of
non-isolated, satellite galaxies have been presented in detail by
\cite{Fattahi2016b,Fattahi2018} for \Apostle{} and by
\cite{Simpson2018} for the \Auriga{} simulations.

Table~\ref{tab:dwarfnums} lists the total number of objects satisfying
these criteria in the various simulation volumes. Given this choice of
mass range and the resolution of \Apostle{} HR and \Auriga{} HR, the
minimum number of particles used to compute DM density profiles is
$\sim 25\,000$, which is more than sufficient to resolve accurately
the dynamics of the inner part of the DM halo, which is the scale of
interest. When we refer to stellar mass, $M_\star$, of a galaxy, we
include all star particles identified by \Subfind{} as being
gravitationally bound. Finally, we exclude any objects that may be
contaminated by the presence of heavier, low-resolution DM particles
-- this is often the case for haloes located too close to the boundary
of the high-resolution region of the simulation volume. This is
achieved by restricting our selection to only dwarfs that are located
within a sphere of radius 1 Mpc from the centre of the main galaxy in
\Auriga{} (3 Mpc from the Local Group barycentre in the case of
\Apostle{}). We have also checked explicitly that no low-resolution
particles are associated with haloes included in the final selection.

To match the isolated haloes between the DMO and hydrodynamical runs,
we use a bijective matching procedure: first, we consider the 50
most-bound DM particles from a candidate halo in the hydrodynamical
run, and look for the DMO halo in which there are at least 25 (50 per
cent) of these particles. The match is then confirmed by repeating the
same process, this time starting with the DMO haloes.

\begin{figure*} \centering \includegraphics[width=\textwidth,trim={7mm 0 0 0}]{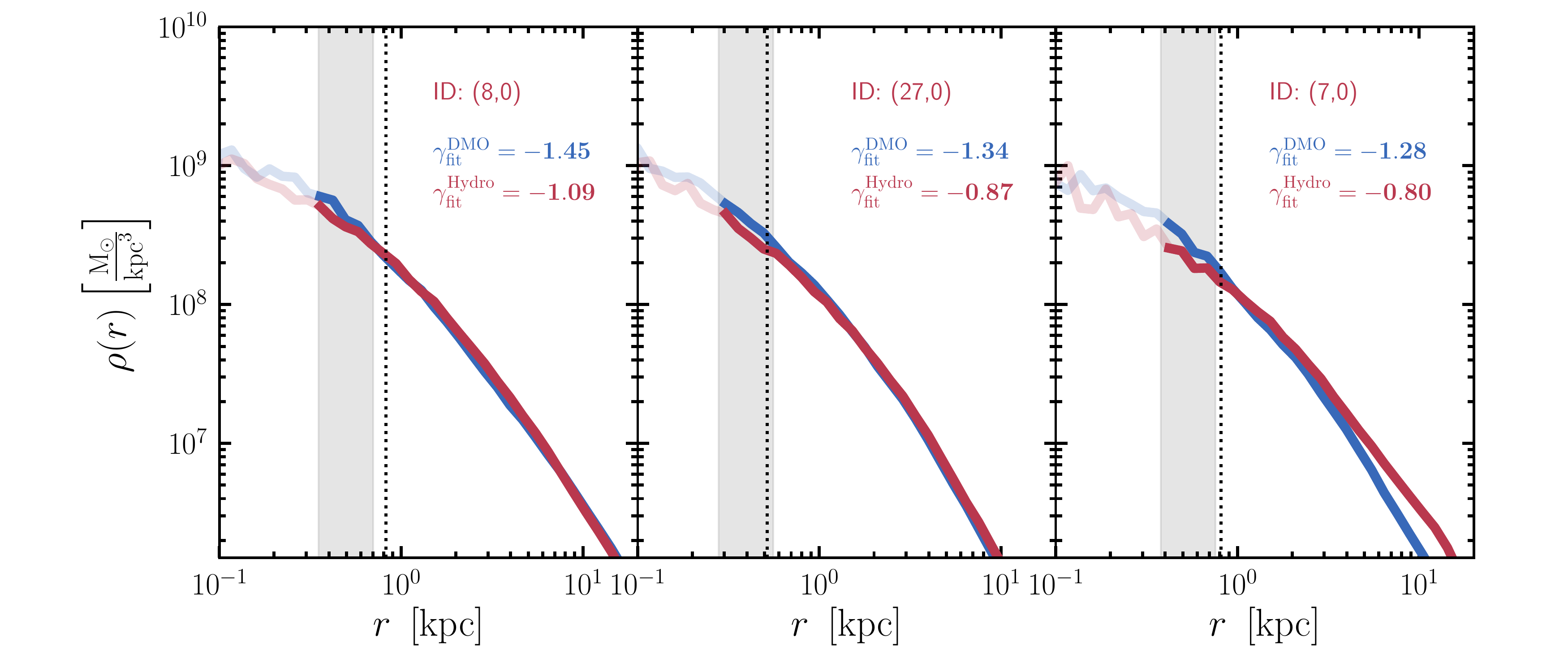}
  \caption{Dark matter density profiles of the isolated dwarf galaxy
    halo that exhibits the shallowest inner slope, $\gamma_{{\rm
        fit}}$, in each of the three hydrodynamical \Apostle{} HR runs
    at $z=0$ (V1, V4 and V6 from left to right). In each panel, the
    thick red line shows the density profile of the dark matter
    component in the run with full hydrodynamics and the thick blue
    line the density profile of this halo's counterpart in the DMO
    version of this simulation. Linestyles are drawn faint below the
    convergence radius of the halo. The vertical dotted line marks 1
    per cent of the halo virial radius. The values of $\gamma_{{\rm
        fit}}$ (as defined in the main text) in the DMO and
    hydrodynamical versions of this halo are compared in the top right
    corner of each panel; the portion of the profile that is fit to
    derive $\gamma_{{\rm fit}}$ is highlighted by the shaded grey
    band. Properties of these dwarfs are listed in
    Table~\ref{tab:dwarfstats}.}
\label{fig:flattestAPOSTLE}
\end{figure*}

\begin{figure*} \centering \includegraphics[width=\textwidth]{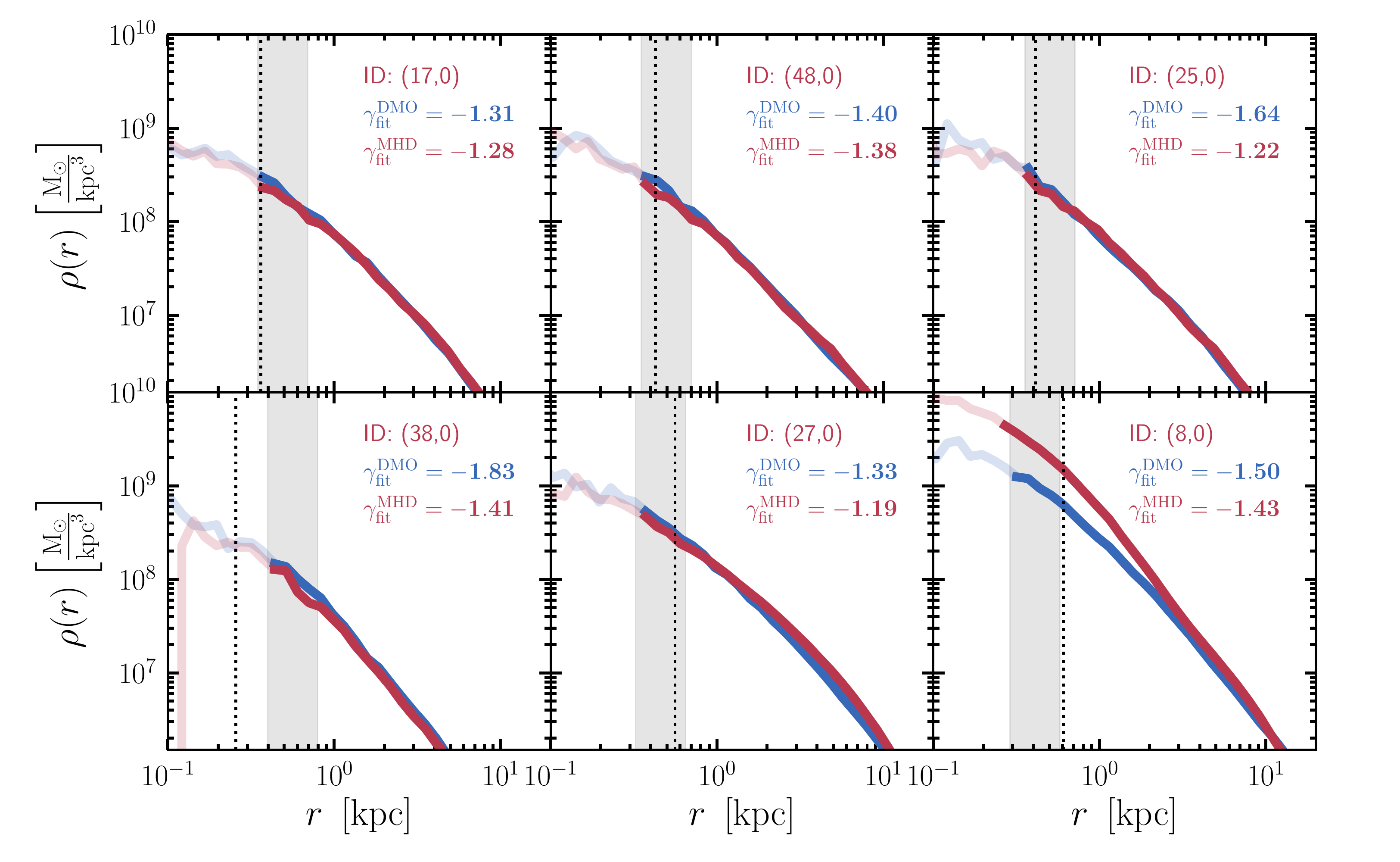}
\caption{Same as Fig.~\ref{fig:flattestAPOSTLE}, but for the six
  \Auriga{} HR haloes (Au-6, 16, 21, 23, 24 and 27 from left to right
  starting from the upper left panel).}
\label{fig:flattestAuriga}
\end{figure*}

\begin{table*}
\centering
\vspace{10pt}
\begin{tabular}{ccccc}
\hline \hline \\
{\bf Simulation} & {\bf (Volume, FOF, Subhalo ID)} & $M_{{\rm DM}}^{{\rm DMO}}$  & $M_{{\rm DM}}^{{\rm Hydro}}$ & $M_{\star}/M_{{\rm DM}}^{{\rm Hydro}}$  \\
           &        & $[{\rm M}_\odot]$ &$[{\rm M}_\odot]$  &   \\
\hline \\
\Apostle{}: & (1)  & (2)  & (3) \\
\hline
     & (Ap-V1, 8,0) & $6.5\times10^{10}$ & $5.3\times10^{10}$ & 0.014  \\
     & (Ap-V1, 38, 0) & $1\times10^{10}$ & $8\times10^9$ & 0.0084  \\ 
    	& (Ap-V4, 22,0) & $1.8\times10^{10}$ & $1.5\times10^{10}$ & 0.0039  \\
      & (Ap-V4, 27,0) & $1.6\times10^{10}$ & $1.4\times10^{10}$ & 0.0039  \\
      & (Ap-V6, 7,0) & $6.2\times10^{10}$ & $7.2\times10^{10}$ & 0.0045  \\
\hline
\Auriga{}: &  &  & &  \\
\hline
  & (Au-6, 17,0) & $5.4\times10^{9}$ & $4.5\times10^{9}$ & 0.0044   \\
  & (Au-16, 47,0) & $7.6\times10^{9}$ & $6.6\times10^{9}$ & 0.0042   \\
 & (Au-16, 48,0) & $8.9\times10^{9}$ & $7.3\times10^{9}$ & 0.0021   \\
 & (Au-21, 25,0) & $8.2\times10^{9}$ & $6.7\times10^{9}$ & 0.002   \\
& (Au-21, 32,0) & $4.7\times10^{9}$ & $3.6\times10^{9}$ & 0.0018   \\
& (Au-23, 15,0) & $8.1\times10^{9}$ & $6.8\times10^{9}$ & 0.007   \\
& (Au-23, 38,0) & $1.9\times10^{9}$ & $1.4\times10^{9}$ & $7.8\times10^{-6}$   \\
& (Au-24, 27,0) & $2.0\times10^{10}$ & $1.8\times10^{10}$ & 0.017   \\
& (Au-24, 52,0) & $9.8\times10^{9}$ & $8.3\times10^{9}$ & 0.011   \\
 & (Au-27, 8,0) & $2.6\times10^{10}$ & $2.2\times10^{10}$ & 0.098   \\
  & (Au-27, 19,0) & $8.8\times10^{9}$ & $7.4\times10^{9}$ & 0.0042   \\
\\ \hline \hline
\end{tabular}
\caption{A list of properties for isolated dwarfs that are given
  individual attention in this paper. A dwarf is identified uniquely
  using the numbers in parentheses provided in column (1), which
  follows the format: (Volume \#, FOF \#, subhalo \#). Column 2 lists
  the mass in DM contained in the DMO counterpart of this halo, while
  column 3 lists the equivalent value in the run with
  hydrodynamics. Finally, column 3 lists the stellar-to-halo mass
  ratio for each dwarf.}
\label{tab:dwarfstats}
\end{table*}

An important characteristic of this work is that while both \Apostle{}
and \Auriga{} are re-simulations of `special' environments, (1) they
are fully cosmological in nature (i.e. the large-scale tidal fields
appropriate to the 100 Mpc volumes they were extracted from are
self-consistently followed albeit at lower resolution), and (2) the
subgrid prescriptions have been shown to produce realistic galaxy
populations (i.e. in agreement with a wide range of observational
data, across a range of redshifts) in larger simulation volumes. Point
(2) in particular is not trivial: for example, a zoom simulation in
which the subgrid parameters are tuned to reproduce properties of
dwarf galaxies on Local Group scales is not guaranteed to reproduce
the galaxy stellar mass function, colour distribution, galaxy
size-mass relation etc. observed among galaxies in the field. The
subgrid models used in \Apostle{} and \Auriga{} are very similar to
those used by the \Eagle{} \citep{Schaye2015} and \Illustris{}
\citep{Vogelsberger2014} simulations, respectively; the galaxy
formation models have not been tuned specifically to reproduce
properties of the Milky Way or galaxies in the Local Group.

To demonstrate that the reverse is also true (i.e. that the chosen
subgrid parameters are appropriate for the resolution / regime of
interest in this paper), in Fig.~\ref{fig:galaxySizes} we present the
galaxy size-stellar mass relation for isolated dwarfs in \Apostle{}
\citep[see also][]{Campbell2017} and \Auriga{}. Galaxy size in this
plot is the (3D) stellar half-mass radius, $r_{1/2}^\star$, while the
stellar mass is the total mass in star particles bound to the halo. We
only display the relations for isolated dwarf galaxies using the
criteria set out at the start of this subsection. For comparison, the
grey diamonds with error bars show the data for isolated dwarf
galaxies in and around the Local Group compiled by
\cite{Mcconnachie2012}. We additionally include data from the SPARC
galaxy sample \citep{Lelli2016} shown in the grey
stars. \cite{Mcconnachie2012} measures the half-light radius along the
semi-major axis of each galaxy, while the values measured in the
simulations are spherical calculations based on the 3D distribution of
stars. To aid the comparison between simulated dwarfs and the data, we
have converted the observed projected half-light radius into the
equivalent 3D half-light radius by multiplying by a factor of
4/3. While the simulations reproduce the general trend seen in the
data, they do not reproduce the scatter at fixed stellar
mass. However, the level of agreement between our simulations and the
data is comparable to that observed in other hydrodynamical
simulations of dwarf galaxies \citep[see, e.g. Fig. 1
  in][]{Fitts2017}. Both \Auriga{} and \Apostle{} simulations show a
paucity of small, compact galaxies ($r_{1/2}^\star < 400$ pc) in the
range $10^6 {\rm M}_\odot < M_\star < 10^8 {\rm M}_\odot$. However,
these sizes are smaller than the minimum resolution with which we are
able to measure density profiles in this work (the `convergence
radius' of the halo; see Section~\ref{sect:DMprof}); as such, the
absence of these galaxies is not expected to impact the remainder of
our analysis in any significant way.

\section{Results}
\label{sect:Results}

In this section, we present the main results of this work. In
particular, we compare the DM density profiles
(Section~\ref{sect:DMprof}) and star formation histories
(Section~\ref{sect:SFHistories}) of isolated dwarf galaxies (using the
criteria outlined in Section~\ref{sect:Definitions}) identified in the
\Apostle{} and \Auriga{} simulations. 

\subsection{The ubiquitous cuspy density profile}
\label{sect:DMprof}

We begin by analysing the shape of DM density profiles of isolated
dwarfs in the \Apostle{} and \Auriga{} simulations.
Fig.~\ref{fig:flattestAPOSTLE} shows the density profiles of the dwarf
galaxy haloes exhibiting the {\it shallowest} inner slope in
\ApostleHydro{} at $z=0$. The inner slope is quantified by a
parameter, $\gamma_{{\rm fit}}$, which is the power law index that
best fits the density profile in the range $r_{{\rm conv}} < r < 2.0
\, r_{{\rm conv}}$, where $r$ is the radial distance from the halo
centre, while $r_{{\rm conv}}$ is the convergence radius defined
according to \cite{Power2003}, and is the radius within which the
relaxation time is $\sim 1/3$ the age of the Universe. This is similar
to the procedure followed by
e.g. \cite{Chan2015,ElBadry2017,Maccio2017}, although these authors
typically fit the range between 1-2 per cent of the halo virial
radius. Our choice of $r_{{\rm conv}}$ is motivated by the fact that
this is the innermost radius of the DM density profile that is
numerically well converged given the number of particles in the halo
-- the profiles shown in Fig.~\ref{fig:flattestAPOSTLE} are drawn with
faint lines below this limit. This figure also shows that the scale
corresponding to 1 per cent the halo's virial radius (vertical dotted
lines) is sometimes located below $r_{{\rm conv}}$, and at other times
does not probe the innermost (resolved) part of the profile, further
motivating our choice to define $\gamma_{{\rm fit}}$ in a range
defined by $r_{{\rm conv}}$. In each panel, the thick red line
represents the DM density profile in \ApostleHydro{}, while the thick
blue curve is the density profile measured for this halo's counterpart
in \ApostleDMO{}.

\begin{figure*}\centering \includegraphics[trim={0 2.7in 0 0},width=\textwidth]{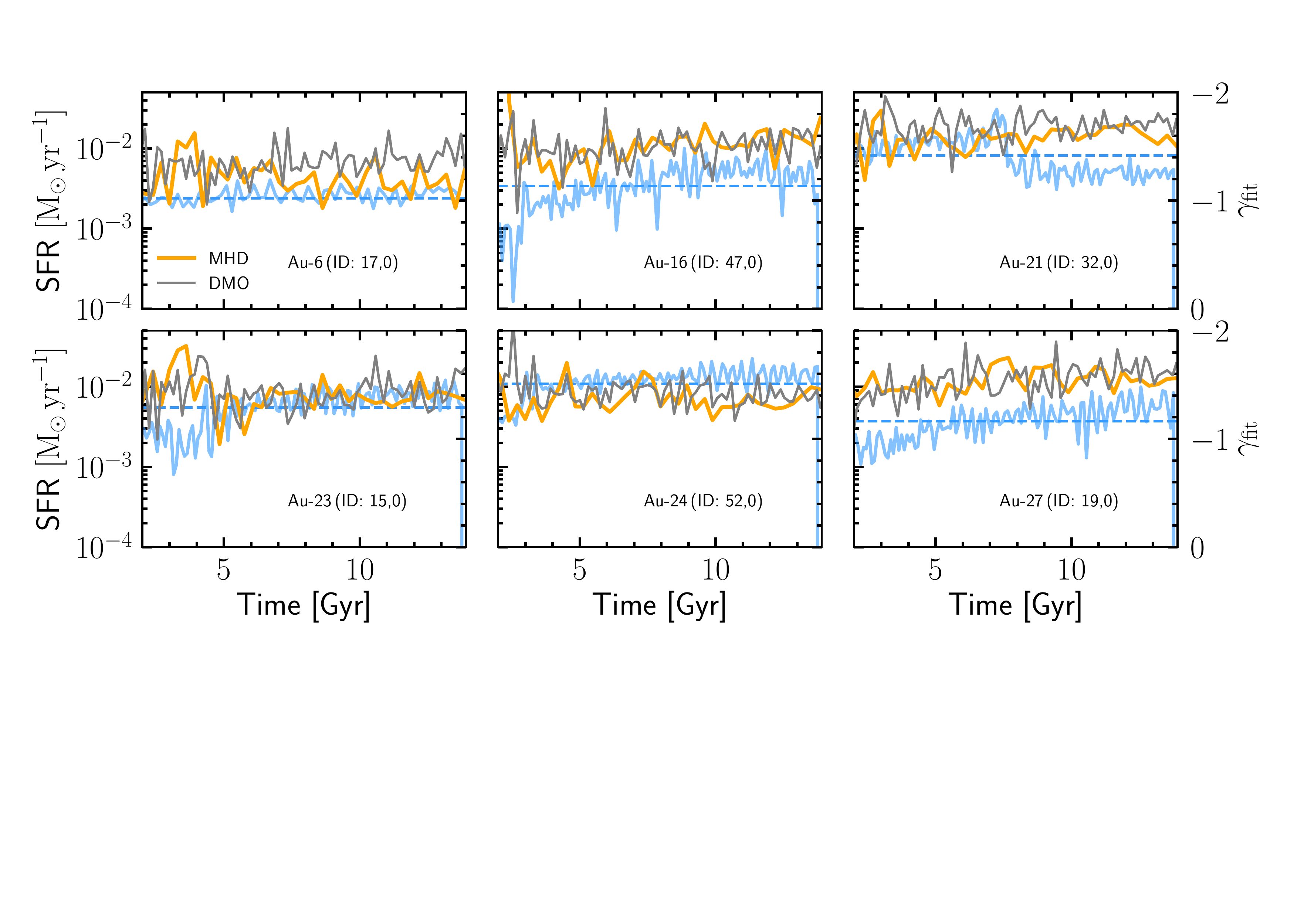}
  \caption{Time evolution of the best-fit inner slope, $\gamma_{{\rm
        fit}}$, of the dark matter density profile in the
    hydrodynamical version of an isolated dwarf galaxy halo (orange),
    and its DMO counterpart (grey) identified in each of the six
    \Auriga{} volumes. The blue curve shows the time variation of the
    star formation rate (smoothed over 100 Myr) of the galaxy formed
    in this halo. The horizontal blue dashed line marks the mean star
    formation rate averaged over the entire history of this galaxy. In
    each panel, we have chosen to display these relations for the
    isolated dwarf galaxy with the greatest stellar mass at $z=0$
    i.e. the halo with the highest {\it average} star formation rate
    in each simulation. Properties of these dwarfs are listed in
    Table~\ref{tab:dwarfstats}.}
\label{fig:SFRgamma}
\end{figure*}

\begin{figure*}\centering \includegraphics[trim={0 0 0 0},width=\textwidth]{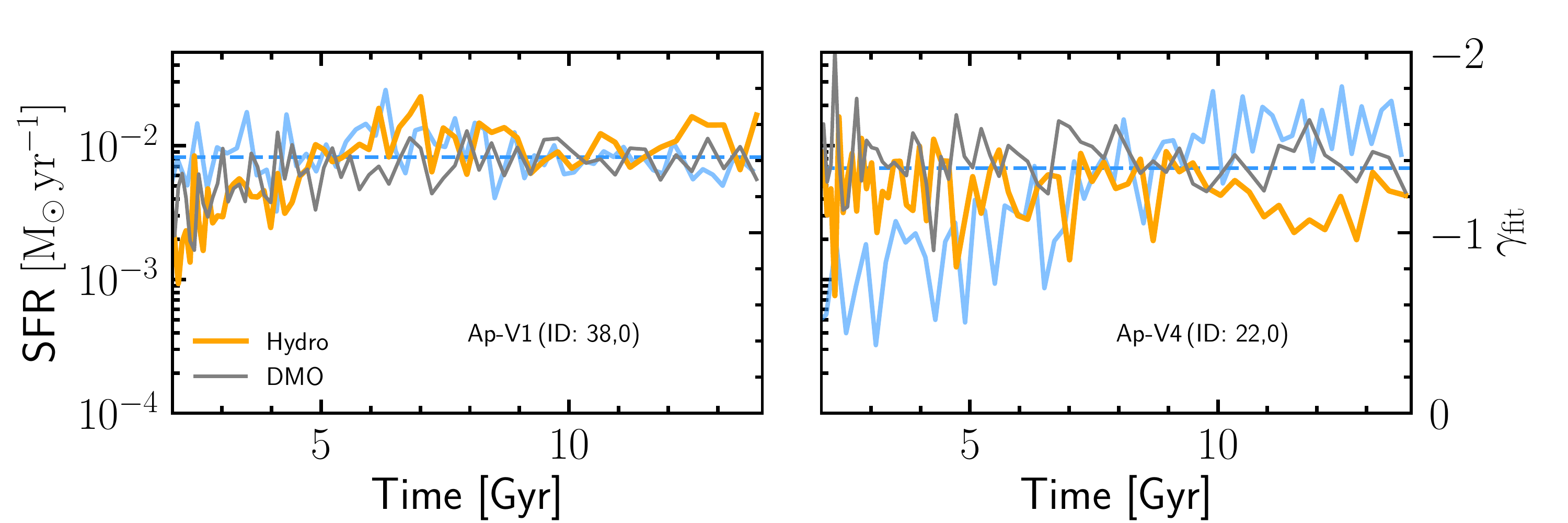}
  \caption{As Fig.~\ref{fig:SFRgamma} for \Apostle{} volumes Ap-V1 and
    Ap-V4.}
\label{fig:SFRgammaAp}
\end{figure*}

Fig.~\ref{fig:flattestAPOSTLE} shows that, according to the values of
$r_{{\rm conv}}$, the DM density profiles of \Apostle{} are reliable
for $r \gtrsim 400$ pc. As expected, our selection of the shallowest
\ApostleHydro{} density profiles yields systems with slightly lower
central densities than in \ApostleDMO{} (within $\lesssim 1$
kpc). However, even the profiles with the shallowest slopes in
\ApostleHydro{} show no evidence of cores, at least larger than 400 pc
in size. In fact, the shallowest slope we measure is $\gamma_{{\rm
    fit}} = -0.80$, associated with a $7.2 \times 10^{10} {\rm
  M}_\odot$ halo in Ap-V4 (right panel of
Fig.~\ref{fig:flattestAPOSTLE}).

The shallowest profiles from \AurigaMHD{} are shown in
Fig.~\ref{fig:flattestAuriga}. Convergence in the density profiles is
reached at a comparable radial scale as in \Apostle{}. While the
central densities are reduced in the runs with MHD relative to DMO
(with the exception of the dwarf galaxy selected from Au-27, shown in
the bottom right panel of Fig.~\ref{fig:flattestAuriga}), once again,
no cores are present. Table~\ref{tab:dwarfstats} lists the properties
of these dwarfs in both simulations. It is interesting to note that
the dwarf galaxy haloes with the shallowest DM density profiles
display a wide range of star formation efficiencies, as measured by
their stellar-to-halo mass ratio, $M_\star/M_{{\rm DM}}$, which ranges
from $8 \times 10^{-6}$ in Au-23 to $\sim 1.5 \times 10^{-2}$ in Ap-V1
and Au-24.

\begin{figure*} \centering
\includegraphics[width=0.48\textwidth]{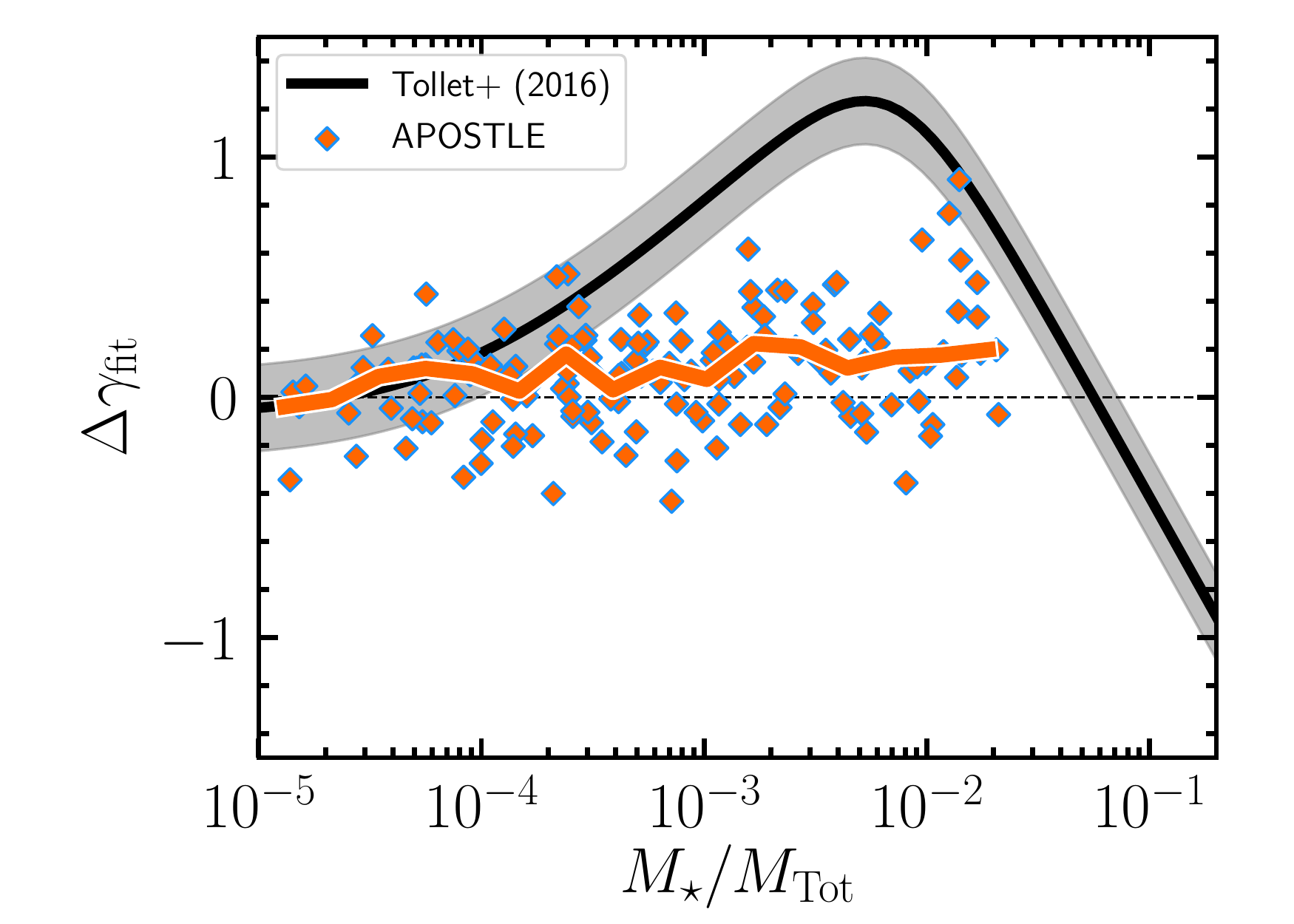}
\includegraphics[width=0.48\textwidth]{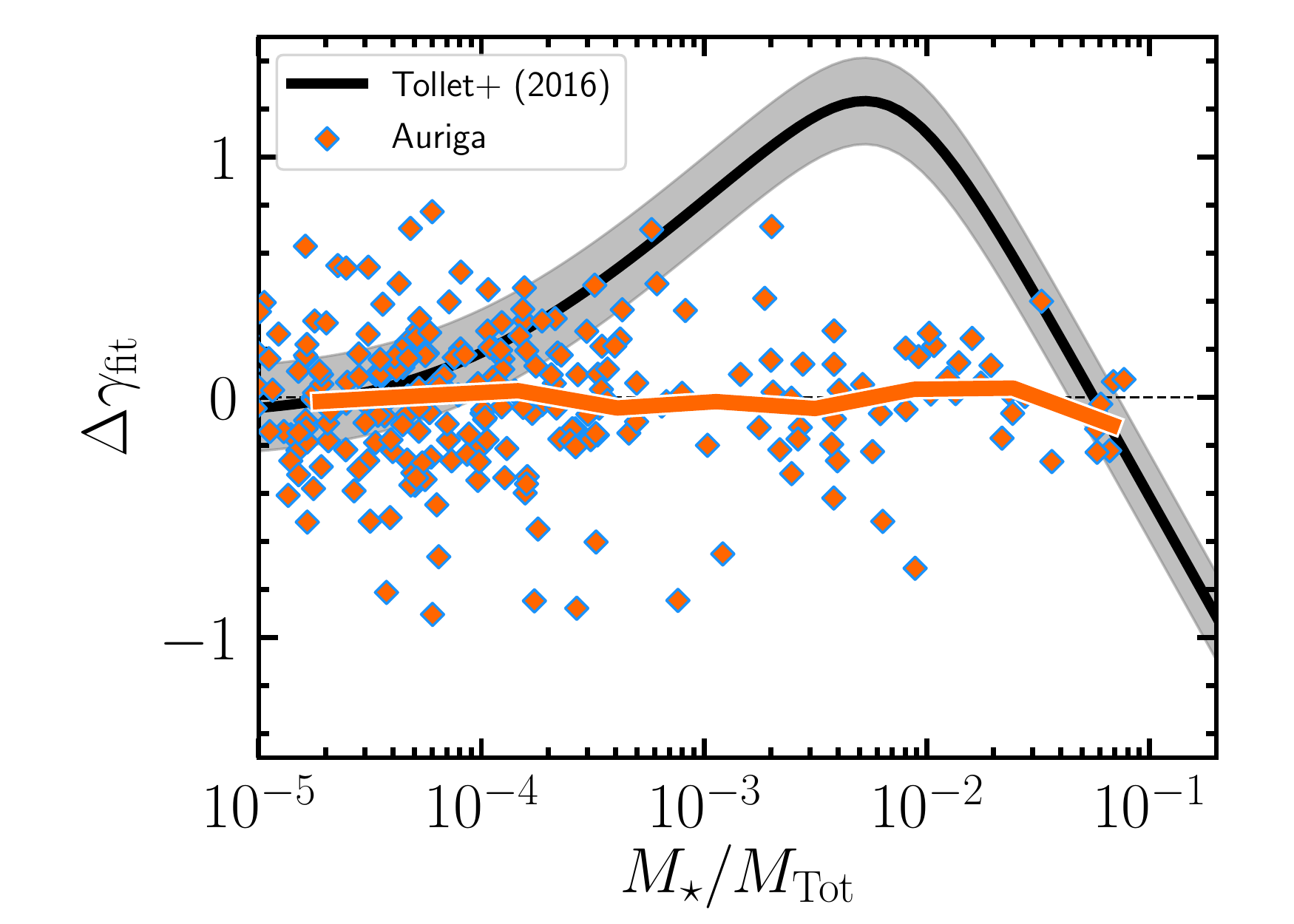}
\caption{Change in the best-fit inner slope of the dark matter density
  profile, $\gamma_{{\rm fit}}$, between isolated \Apostle{} (left
  panel) and \Auriga{} (right panel) haloes and their matched DMO
  counterparts, as a function of stellar-to-total halo mass
  ratio. Negative values correspond to steeper dark matter density
  profiles in the hydrodynamical runs, while positive values
  correspond to profiles that have become shallower in the
  hydrodynamical runs. Each diamond represents an individual halo,
  while the solid line shows the median relation. The solid black
  curve is obtained using the fitting function proposed by
  \citet{Tollet2016}, building on a similar relation previously
  suggested by \citet{DiCintio2014}; here we have assumed
  $\gamma_{{\rm fit}}^{{\rm DMO}} = -1.5$.} The grey band represents a
1$\sigma$ scatter of 0.18 around the mean relation.
\label{fig:deltaGammaAPOSTLE}
\end{figure*}

\subsection{Cusps and bursty star formation}

As discussed in Section~\ref{sect:Intro}, core formation in the
literature has been ascribed to energetic processes associated with
galaxy formation, such as repeated outbursts of supernovae, and the
existence of bursty and sustained periods of high star formation rates
(SFRs). A particularly interesting connection between SFRs and the
shape of the DM density profile was demonstrated by
\cite{ElBadry2017}, who found a strong anti-correlation between the
two in high-resolution simulations of dwarf galaxies; where periods of
bursts in the SFR were associated with a flattening of $\gamma_{{\rm
    fit}}$, whereas a steeper value of $\gamma_{{\rm fit}}$ was
restored during more quiescent phases. Simulations performed by
\cite{Read2016b} also find differences in the rotation curve of dwarf
galaxies induced by episodes of starbusts and quiescence.

To examine if such a correlation can be identified in our simulations,
in Fig.~\ref{fig:SFRgamma} we plot the time evolution of $\gamma_{{\rm
    fit}}$ for a selection of isolated dwarfs from \AurigaDMO{} (grey
curves) and \AurigaMHD{} (orange curves), and their associated SFRs
(blue curves). We have specially selected isolated dwarfs from
\AurigaMHD{} that have the highest stellar mass at $z=0$. While the
merger tree of a galaxy can be traversed to trace the growth of
stellar mass and measure the SFR, the resolution of this method is
limited by the spacing of simulation snapshots. On the other hand, the
age of a stellar population is output at the exact timestep
corresponding to its birth. This means that for all stars identified
in a galaxy at a particular time, the snapshots contain information on
the exact scale factor at which this star was born; this information
can be used to create a star formation history (SFH) with as good a
time resolution as it is possible to obtain from the simulations. In
what follows, we always measure SFRs/SFHs using the latter
definition. In Fig.~\ref{fig:SFRgamma}, the SFR of each galaxy has
been smoothed over a 100 Myr interval.

The specific SFRs for our selection of \AurigaMHD{} dwarfs are
comparable (and, in some cases, larger) than those reported by
\cite{Fitts2017} and \cite{ElBadry2017}. From Fig.~\ref{fig:SFRgamma},
we find that in no case does the value of $\gamma_{{\rm fit}}$ ever
become shallower than $\approx -1$; in fact, the evolution of
$\gamma_{{\rm fit}}$ is largely identical in \AurigaMHD{} and
\AurigaDMO{}. In other words, the effect of the hydrodynamics, if any,
on the shape of the DM density profile is {\it comparable to the
  natural variation of the inner slope} (due to mergers and accretion)
that one measures from a purely collisionless
simulation. Fig.~\ref{fig:SFRgamma} therefore shows that in the six
\AurigaMHD{} simulations, even transient cores (i.e. those that form
temporarily, before reverting to a cusp) never form. As shown in
Fig.~\ref{fig:SFRgammaAp}, we find similar results for haloes in the
\Apostle{} simulation.

\subsection{Cusps and galaxy formation efficiency}

Several authors have reported a positive correlation between the value
of the inner slope of the DM density profile (i.e. $\gamma_{{\rm
    fit}}$) and the star forming efficiency of a halo, measured by its
stellar-to-halo mass ratio
\citep[e.g.][]{Governato2012,DiCintio2014,Maccio2017}. In principle,
such a suggestion is reasonable: if star formation and/or feedback is
responsible for altering the shape of the DM density profile, haloes
with larger stellar-to-halo mass ratio are more likely to be affected
as there is effectively more energy available from supernovae to
unbind the dark matter. Furthermore, \cite{Fitts2017} find that, in
their simulations, the half-mass radius of the galaxy sets a
characteristic length scale which determines the size of the core
formed in the DM density profile.

Fig.~\ref{fig:deltaGammaAPOSTLE} investigates the relationship between
$\gamma_{{\rm fit}}$ and $M_\star/M_{{\rm Tot}}$ (where $M_{{\rm
    Tot}}$ is the total halo mass including DM, gas and stars) in
\Apostle{} and \Auriga{}. Rather than simply plotting $\gamma_{{\rm
    fit}}$ from the hydrodynamical run on the vertical axis (as is
commonly done in the literature), we plot $\Delta \gamma_{{\rm
    fit}}~=~\gamma_{{\rm fit}}^{{\rm Hydro}}~ -~\gamma_{{\rm
    fit}}^{{\rm DMO}}$ i.e. the {\it change} in the inner slope
between a matched pair of hydro / DMO haloes. The reason for this is
that smaller haloes, which are less well resolved, will naturally
yield more `negative' values of $\gamma_{{\rm fit}}$ as $r_{{\rm
    conv}}$ in these haloes will be closer to the scale radius of the
profile, where the slope $\approx -2$. For larger haloes, which are
resolved with many more particles, $r_{{\rm conv}}$ is pushed `further
in' towards the halo centre where the typical slope is closer to
$\approx -1$. As low-mass haloes, on average, have low star forming
efficiency, one would measure a positive correlation between
$\gamma_{{\rm fit}}$ and $M_\star/M_{{\rm Tot}}$ that is in reality is
just an artefact. As defined, {\it negative} values of $\Delta
\gamma_{{\rm fit}}$ correspond to profiles that have become {\it
  steeper} in the simulation with hydrodynamics, while {\it positive}
values of $\Delta \gamma_{{\rm fit}}$ correspond to haloes where the
slope is {\it shallower} after the inclusion of baryons.

\begin{figure*} \centering
\includegraphics[width=\textwidth,trim={0.4in 0.4in 0 0}]{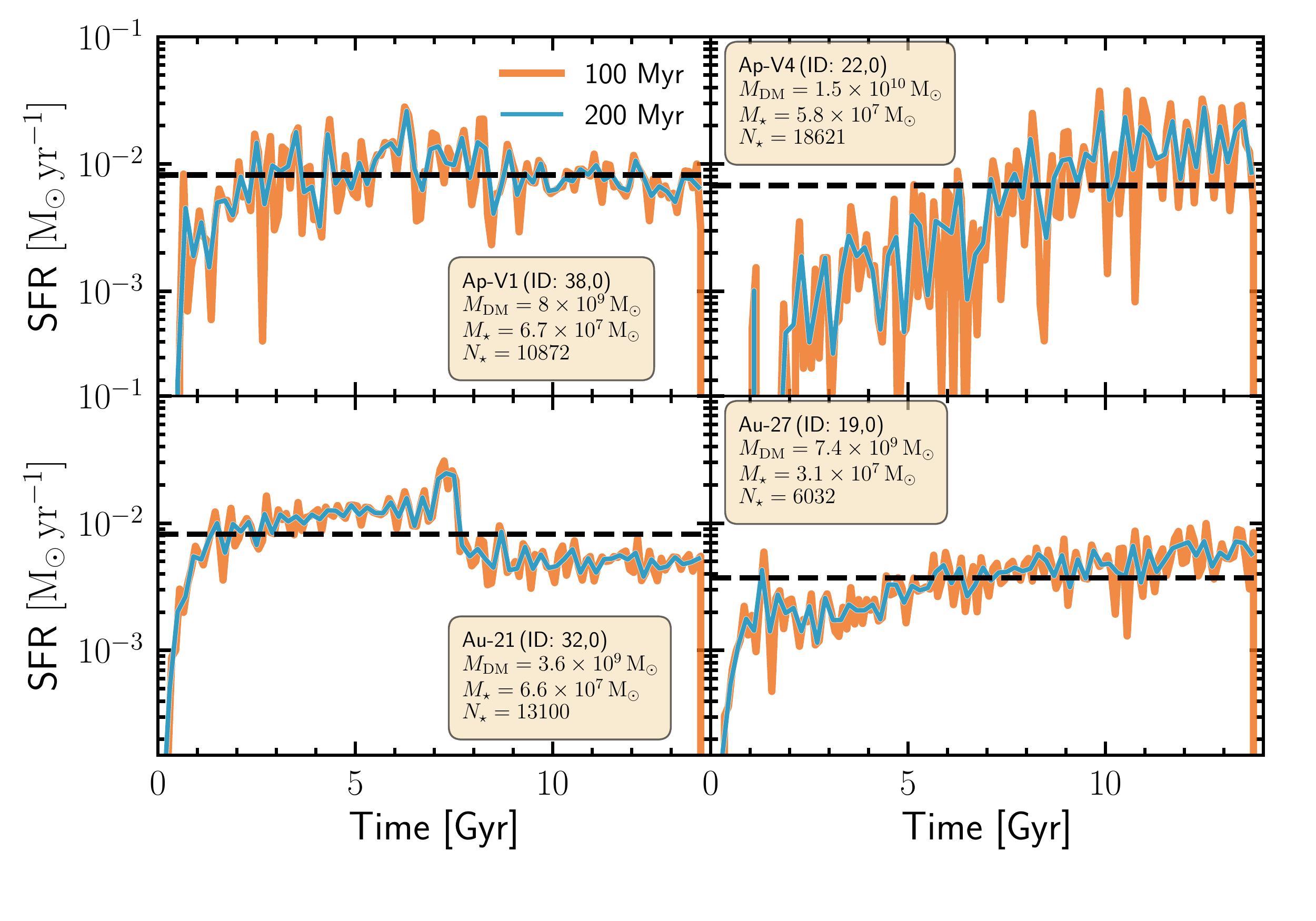}
\caption{Individual star formation histories (SFHs) for a selection of
  isolated dwarf galaxy haloes from \Apostle{} (top row) and \Auriga{}
  (bottom row). These galaxies were selected to have the highest
  average SFR amongst all isolated dwarfs at $z=0$ in the volume from
  which they are chosen. After collecting the set of stars present in
  each galaxy at $z=0$, the expansion factor at which the star
  particle was born is used to construct the SFH. The orange and blue
  lines respectively show the SFHs smoothed over 100 and 200 Myr time
  intervals. The dashed horizontal line marks the average star
  formation rate of the galaxy in each panel.}
\label{fig:SFR}
\end{figure*}

The orange lines in Fig.~\ref{fig:deltaGammaAPOSTLE} show the median
trend. Given the relatively small number of isolated dwarfs in the two
simulations and the scatter in $\Delta \gamma_{{\rm fit}}$, the median
trend is noisy. However, there is no obvious trend of $\Delta
\gamma_{{\rm fit}}$ with $M_\star/M_{{\rm Tot}}$; the variations are
consistent with zero. For comparison, we have also included the
relationship inferred from simulations by \cite{DiCintio2014} and
\cite{Tollet2016}, which shows a clear variation in $\Delta
\gamma_{{\rm fit}}$ as a function of $M_\star/M_{{\rm Tot}}$. In
making this comparison with \cite{Tollet2016}, we have assumed
$\gamma_{{\rm fit}}^{{\rm DMO}} = -1.5$.

\begin{figure*} \centering
\includegraphics[scale=0.6,trim={0.5in 0 0 0}]{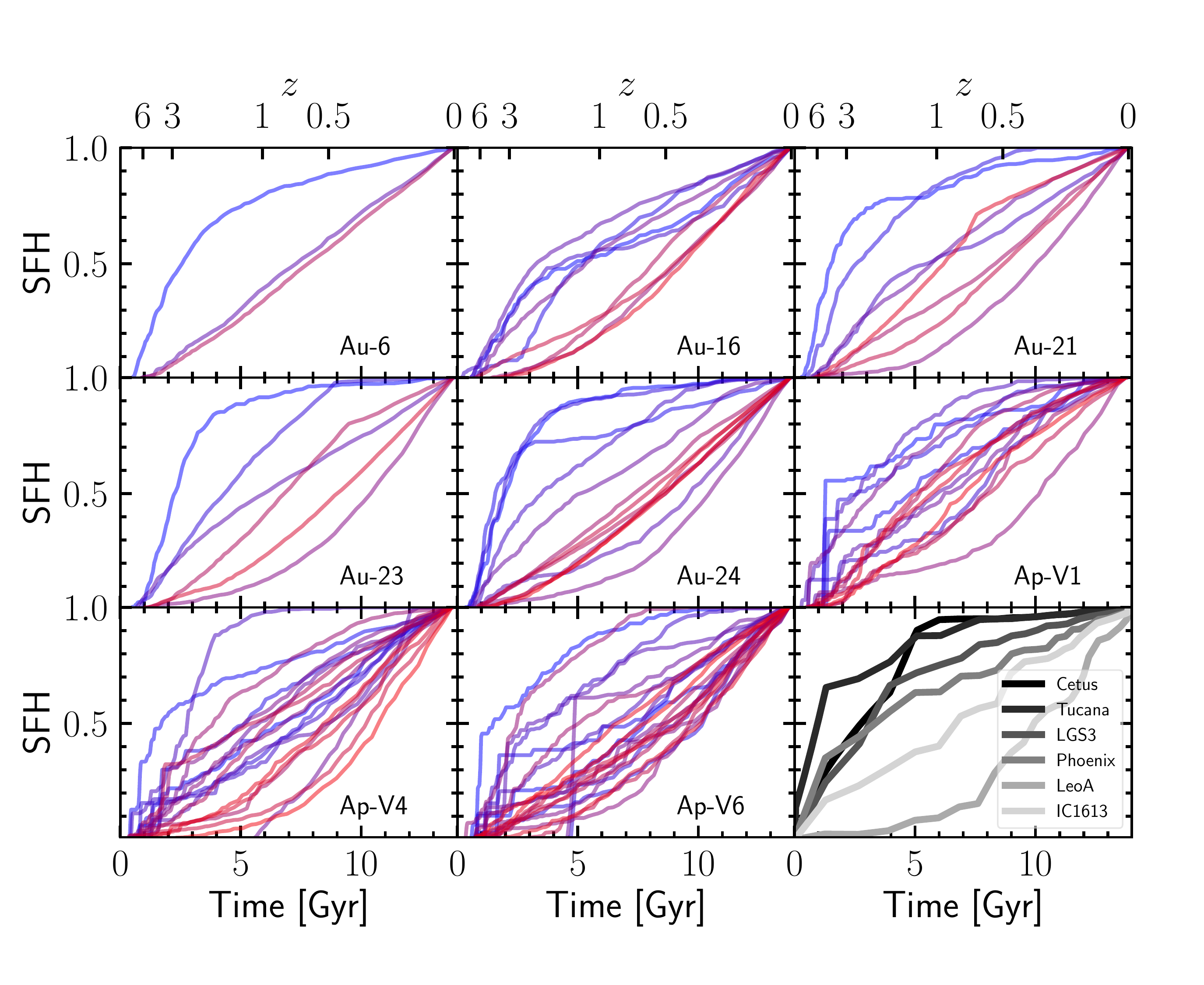}
\caption{Cumulative SFHs for all isolated dwarf galaxies in the mass
  range $10^9 < M_{{\rm DM}} / {\rm M}_\odot < 5 \times 10^{10}$ and
  $10^6 < M_\star / {\rm M}_\odot < 10^8$ in the \Auriga{} (panels
  1-5) and \Apostle{} HR runs (panels 6-8). As in Fig.~\ref{fig:SFR},
  the SFHs are constructed using the stellar birth time of star
  particles identified at $z=0$ in each galaxy. The colour of each
  line, from blue to red, ranks galaxies in ascending order of present
  day stellar mass. To compare with the SFHs from our simulations, the
  last panel also displays the SFHs measured for real dwarf galaxies
  by \citet{Skillman2014}.}
\label{fig:sfhAPOSTLE}
\end{figure*}

The density profiles shown in Figs.~\ref{fig:flattestAPOSTLE}
and~\ref{fig:flattestAuriga} show no significant deviation from an NFW
shape, and lack a characteristic length scale that may be imposed by
the galaxy half-mass radius ($r_{1/2}^{\star}$) on the host halo DM
profile. We remind the reader that the size-mass relations of isolated
dwarfs in both \Apostle{} and \Auriga{} are consistent with the data
(Fig.~\ref{fig:galaxySizes}). Furthermore, for galaxies with $M_\star
> 10^7 {\rm M}_\odot$, $r_{1/2}^{\star} \gtrsim 600$ pc $\approx 1.5
\, r_{{\rm conv}}$ in both \Apostle{} and \Auriga{}, so any potential
scale imprinted on the DM density profile would have been adequately
resolved in our simulations.

\subsection{Cusps and star formation history diversity}
\label{sect:SFHistories}

Next, we proceed to examine the star formation histories (SFHs) of the
isolated dwarfs identified in our simulations. In Fig.~\ref{fig:SFR},
we show the evolution of SFRs for a selection of individual galaxies
from \ApostleHydro{} (top panel) and \AurigaMHD{} (bottom panel). The
orange and blue lines, respectively, show the SFRs averaged over 100
and 200 Myr time bins. We have chosen isolated dwarf galaxies that
have the largest $z=0$ stellar mass in the volume from which they are
extracted. It is interesting to observe the appreciable fluctuations
in the SFRs of these galaxies, particularly in the case of the
\ApostleHydro{} dwarfs. For example, the galaxy selected from Ap-V4
shows fluctuations in SFR of over two orders of magnitude over 100 Myr
intervals. The dwarf galaxies from \AurigaMHD{} also show big temporal
variations in SFR, although these galaxies are not as bursty as those
in \ApostleHydro{}. We have checked explicitly that the burstiness is
not due to stochastic sampling in the star formation prescription:
typically, each time bin in the smoothed SFH contains hundreds of
newly-formed star particles, while the time intervals over which star
formation is averaged are well above the length of a typical timestep
taken in the simulation. 

For objects of similar mass, \cite{Sparre2017} found that galaxies in
the \Fire{} simulations display strong, short bursts of star formation
over 10 Myr timescales. When comparing the SFRs of \Apostle{} and
\Auriga{} galaxies smoothed over 10-50 Myr timescales we find that, in
general, the dwarfs in our simulations exhibit more gentle SFR
fluctuations than in \Fire{}, where galaxies show a stronger
post-burst phase (i.e. a burst of star formation in the last $\sim$
200 Myr or so of evolution). Recently, \cite{Dutton2018} have also
reported larger SFR fluctuations in core-forming dwarfs than that
measured in the cuspy dwarfs from \Apostle{} and \Auriga{}. This is,
in part, due to the different timescales over which the SFR is
averaged: \cite{Dutton2018} average SFR over $\sim$5 Myr, which is
considerably shorter than our choice of 100-200 Myr. Our conservatism
is motivated by the desire to stay clear of the regime in which the
stochastic formation of indvidual star particles may manifest as
burstiness. In any case, we do not believe that this difference in the
degree of SFR burstiness is the reason for the lack of cores in the
simulations we have presented: indeed, as \cite{Llambay2018} have
shown, even extremely bursty dwarfs may continue to exhibit cuspy DM
density profiles \citep[see also][and the discussion in
  Section~\ref{sect:coreFormation}]{Revaz2018}.

It is natural to ask if the fluctuations in the SFR of the \Apostle{}
and \Auriga{} galaxies seen in Fig.~\ref{fig:SFR} are compatible with
the inferred SFHs of dwarfs observed in the Local
Group. Fig.~\ref{fig:sfhAPOSTLE} shows the cumulative SFHs of dwarf
galaxies in \AurigaMHD{} (panels 1-5) and \ApostleHydro{} (panels 6-8)
having stellar mass $10^6 < M_\star/{\rm M}_\odot < 10^8$ at $z=0$;
each curve represents a single galaxy. The final panel in this figure
displays measured SFHs for real dwarf galaxies compiled by
\cite{Skillman2014}, who infer stellar ages by fitting the
colour-magnitude diagrams assuming a stellar population synthesis
model. The selection on stellar mass applied in
Fig.~\ref{fig:sfhAPOSTLE} is consistent with the stellar masses of the
galaxies in the \cite{Skillman2014} dataset.

Dwarf galaxies in both sets of simulations exhibit very diverse
SFHs. The comparatively smaller simulation volume in \Auriga{}
compared to \Apostle{} results in fewer galaxies satisfying our
criteria for isolated dwarfs in the appropriate stellar mass
range. While the majority show sustained stellar mass growth
throughout cosmic time, there are populations of dwarfs that are early
forming (in which, for example, 80 per cent of the mass has been
accumulated by $z=3$) and late forming (more than half of the mass is
accumulated after $z=0.5$). The diverse SFHs are broadly comparable to
those of observed Local Group dwarfs shown in the final panel of
Fig.~\ref{fig:sfhAPOSTLE}.

Another important observation can be made from
Figs.~\ref{fig:SFR}~and~\ref{fig:sfhAPOSTLE}. It is clear from
Fig.~\ref{fig:SFR} that galaxies in \AurigaMHD{} typically have more
quiescent SFHs than galaxies in \ApostleHydro{}. Yet, the {\it
  cumulative} SFHs in both simulations are similar. This demonstrates
that the integrated SFH cannot inform us of whether the {\it
  differential} version of the SFH (as in Fig.~\ref{fig:SFR}) is
bursty or not. Both bursty and comparatively quiescent SFHs can match
the integrated SFHs inferred from the data; however, this agreement
does not reveal which, if any, SFH is more realistic.

\section{Discussion}
\label{sect:coreFormation}

In Section~\ref{sect:DMprof} and~\ref{sect:SFHistories}, we have found
that even though isolated dwarf galaxies in \Apostle{} and \Auriga{}
have bursty SFHs (comparable to those in other papers in the
literature), their DM haloes do not form cores -- at least not with a
size $\gtrsim 400$ pc, which is the nominal resolution (determined by
the convergence radius) at which our density profiles are
reliable. Core formation in hydrodynamical simulations is attributed
to late-time bursts of star formation and the resulting gas motions
that cause fluctuations in the gravitational potential of the DM
\citep[e.g.][]{Pontzen2012}. In this section, we estimate the energy
released by supernovae in our simulations and discuss why cores do not
form in them.

The relevant timescale for inducing lasting changes to the DM density
profile is the dynamical time of the halo at the spatial scale of
interest, $t_{{\rm dyn}}$. We now make an estimate of the energy
released by supernovae in \Apostle{} and \Auriga{} dwarfs over a
dynamical time at $\sim 1$ kpc, which corresponds roughly to the core
size of interest.

Both sets of simulations adopt a Chabrier stellar initial mass
function (IMF). Assuming that only stars with mass 8-100\,${\rm
  M}_\odot$ explode in core-collapse supernovae, and that each
supernova releases $\sim 10^{51}$ ergs of energy, we estimate that
energy of the order of $\sim 2 \times 10^{49}$ ergs$/{\rm M}_\odot$ is
injected per stellar mass in stars formed. Within the dynamical time
at 1 kpc from the halo centre, a galaxy is able to produce at most
$\Delta M_\star = {\rm SFR} \times t_{{\rm dyn}}^{{\rm 1 kpc}}$, where
SFR is the star formation rate of the galaxy during this period. The
total energy available from supernovae is then:
\bq \label{eq:coreform}
E_{{\rm SN}} &=& 2\times10^{49} {\rm ergs} \cdot \Delta M_\star \nonumber \\
 &=& 2\times10^{49} {\rm ergs} \cdot {\rm SFR} \times t_{{\rm dyn}}^{{\rm 1 kpc}},
\eq
where $E_{{\rm SN}}$ is the energy released in supernovae following
the formation of $\Delta M_\star$ in stellar mass.
Inserting typical values for the SFR and $t_{{\rm dyn}}^{{\rm 1 kpc}}$
for $\sim 10^{10} {\rm M}_\odot$ dwarfs in \Auriga{} and \Apostle{},
we obtain $E_{{\rm SN}} \sim \mathcal{O}(10^{55})$ ergs (the precise
value for an individual galaxy will depend on its SFR and the
concentration of its host halo)\footnote{This calculation assumes a
  feedback event that occurs in a single, extended burst.
  \cite{GarrisonKimmel2013} have argued that a single explosive event
  is typically more effective than short, repeated bursts (totalling
  to the same overall outflow mass) at {\it reducing} the central
  densities of DM haloes; on the other hand, multiple cycles of
  outflows are more effective at producing {\it shallower} density
  slopes.}. While only a fraction of this energy will couple to the
DM, the total energy budget available from star formation in these
simulations is consistent with estimates in the literature
\citep[e.g.][]{Pontzen2012,Penarrubia2012,Chan2015}, and of a similar
order of magnitude to the gravitational work needed to unbind a cusp
into a core. This, combined with our findings in
Section~\ref{sect:Results}, demonstrates that bursty SFHs and feedback
from supernovae are {\it not}, by themselves, {\it sufficient}
conditions for forming cores in dwarf galaxy haloes.

\begin{figure} \centering
\includegraphics[width=\columnwidth]{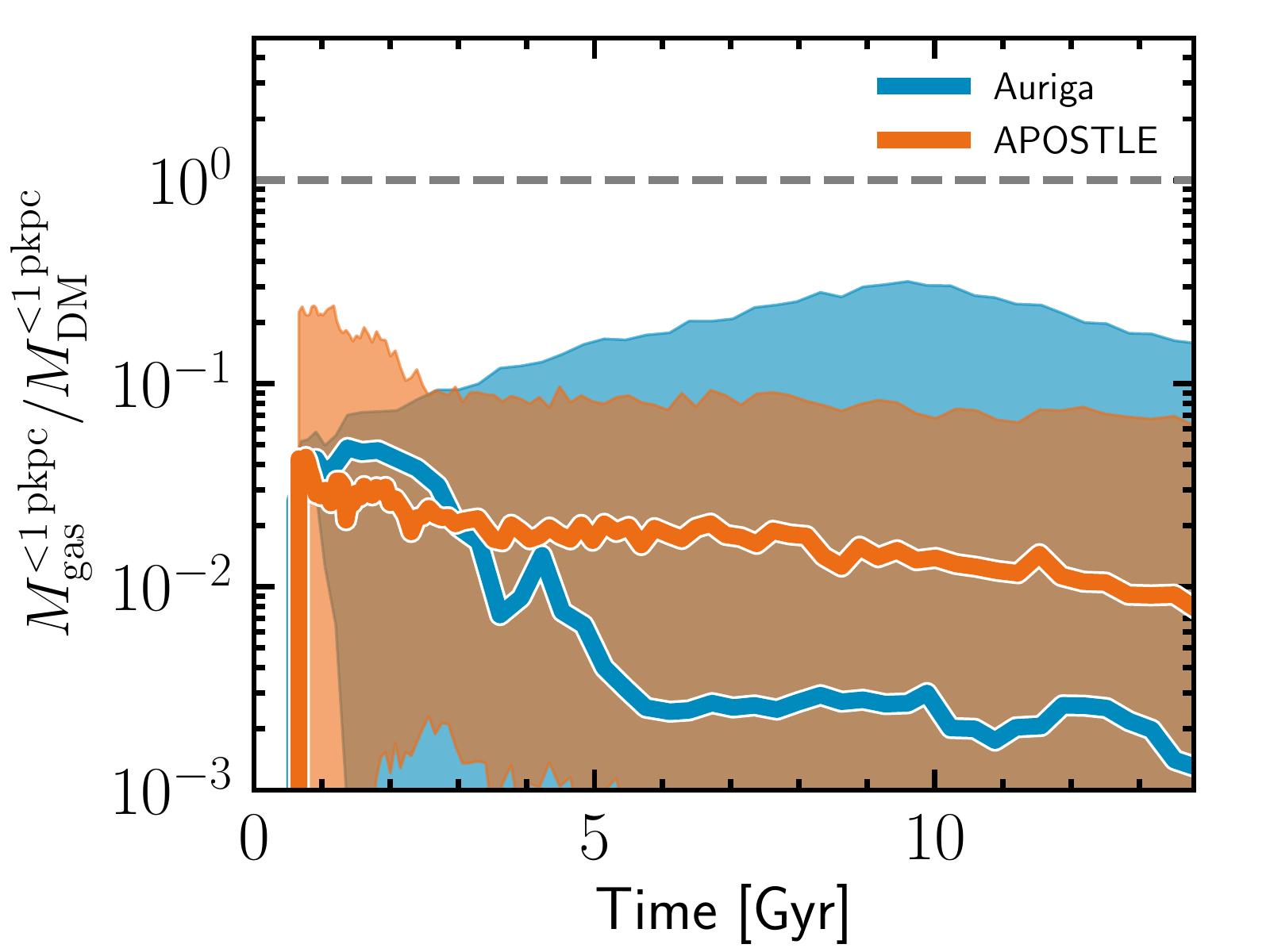}
\caption{Evolution of the ratio of total gas mass to mass in dark
  matter within the central one physical kiloparsec of isolated dwarfs
  in the \Auriga{} (blue) and \Apostle{} (orange) simulations. The
  thick solid lines show the median ratios, while the shaded regions
  encompass the 10$^{{\rm th}}$ and 90$^{{\rm th}}$ percentiles.}
\label{fig:gasDMratio}
\end{figure}

One reason that may explain, at least in part, why both \Auriga{} and
\Apostle{} fail to produce cores can be traced back to the observation
made by \cite{Governato2010} that the core-forming ability of a
simulated dwarf galaxy is also sensitive to the gas density threshold
for star formation, $n_{{\rm sf}}$, assumed in the simulation. The
interpretation is that with a higher star formation threshold, more
gas is allowed to collect at the centre of a DM halo, eventually
resulting in the gas density exceeding the local DM density. When star
formation eventually occurs, the resulting gas outflow in a simulation
with a high threshold is more effective at expanding the orbits of DM
particles near the halo centre, unbinding a fraction of these
particles and eventually leading to the formation of a core, as
proposed originally by \cite{NEF1996}.

In both \Apostle{} and \Auriga{}, $n_{{\rm sf}} = \mathcal{O} (0.1)\,
{\rm cm}^{-3}$.  By contrast, in the works of
\cite{Governato2010,Governato2012,DiCintio2014,Onorbe2015,Fitts2017,Maccio2017}
-- where the formation of cores in dwarf galaxies haloes has been
reported -- the typical values of $n_{{\rm sf}}$ range from 10-1000
cm$^{-3}$, that is between 100 to 10,000 times larger than the value
adopted in \Apostle{} and \Auriga{}. To draw an analogy with the
\cite{NEF1996} mechanism, the gravitational potential of the gas in
these simulations is allowed to build up to much larger values than in
our simulations, with the result that the eventual episodes of
feedback from star formation generate gas motions that are more
effective at perturbing the orbits of neighbouring DM particles. The
absence of cores in \Apostle{} and \Auriga{} is therefore consistent
with the predictions of \cite{Governato2010} who showed that a low
threshold density $\mathcal{O} (0.1)\,{\rm cm}^{-3}$ (as we have
adopted in the present work) is ineffective as forming a core; a value
closer to $\mathcal{O} (100)\,{\rm cm}^{-3}$ is required for gas to
become concentrated enough to dominate the gravitational potential
near the centre.

A consequence of the relatively low threshold for star formation
adopted in our simulations is that gas is converted into stars before
it is allowed to become gravitationally dominant over the DM. This is
demonstrated explicitly in Fig.~\ref{fig:gasDMratio}, which shows the
evolution with time of the ratio of the mass in gas to the mass in DM
within one physical kiloparsec for dwarfs in \Apostle{} and
\Auriga{}. The solid lines represent the median ratios over the age of
the Universe, while the shaded regions encompass the 10$^{{\rm
    th}}$-90$^{{\rm th}}$ percentiles of the population. This figure
shows that total gas mass (and, by extension, gas potential) is always
gravitationally subdominant to the DM for all simulated dwarfs. Any
fluctuations in the potential that may be induced by gas motions
following a feedback event are therefore ineffective at perturbing the
potential of the DM particles over the same physical scale, and these
systems remain DM-dominated at all times. A systematic demonstration
of the effect of varying $n_{{\rm sf}}$ in simulations similar to
\Apostle{} has been presented by \cite{Llambay2018}.

It is important to stress that, in this picture, the entire history of
the gas content in dwarfs is relevant, rather than simply how much
there is at present day. For example, while the dwarfs that are
claimed to have cores may be DM-dominated today, for the core to have
formed in the first place the gas content within the inferred core
size must have been gravitationally-dominant over the DM. After this
gas is eventually expelled by supernovae (potentially forming a core
through induced fluctuations in the local potential), it need not
return. In principle, therefore, it is possible for dwarfs that are
DM-dominated at present to exhibit cores; considering the entire
history of the gas content of these galaxies, which is presently
inaccessible in the data, is key.

Finally, it is worth highlighting that there is still considerable
debate as to how prevalent cores are in observed dwarfs. As we have
discussed previously, there are a number of systematic effects in the
techniques used to infer DM density profiles from observational
data. For example, \cite{Read2016b}, \cite{Oman2017} and
\cite{Pineda2017} have emphasised the importance of accounting for the
presence of thick H\,{\sc i} disks and non-circular motions of gas
when measuring H\,{\sc i} rotation curves. In their mock
`observations' of galaxies from the \Apostle{} project \cite{Oman2017}
find that viewing these galaxies from different lines-of-sight results
in a diverse set of rotation curves {\it for the same galaxy}. In some
cases, particular orientations result in a severe underestimate of the
circular velocity in the inner halo, producing a `core-like' rotation
curve when, in fact, the 3D DM density distribution in the simulation
has a cusp.

The spatial distribution of stellar populations with kinematically
distinct metallicity components in some dwarf galaxies has also been
used to infer the mass profile of the surrounding DM halo
\citep[e.g.][]{Battaglia2008,Amorisco2012,Strigari2014}. Using this
technique, \cite{Walker2011} inferred the existence of cores in both
the Sculptor and Fornax dwarf spheroidal galaxies. However, as shown
recently by \cite{Genina2017}, using galaxies extracted from
\Apostle{}, even this method is sensitive to the viewing angle used to
measure the kinematics of these metallicity populations; in
particular, the assumption of spherical symmetry can mistakenly lead
to the inference of a core when there is actually a cusp.

\section{Conclusions}
\label{sect:Conclusions}

We have carried out a detailed investigation of the dark matter (DM)
density profiles of isolated dwarf galaxy haloes in the
high-resolution \Apostle{} and \Auriga{} cosmological, hydrodynamical
simulations. We have focused specifically on their inner profiles in
the context of claims that the presence of cores inferred from the
rotation curves of some observed dwarf galaxies represents a
shortcoming of the popular cold dark matter (CDM) paradigm, wherein
collisionless DM-only simulations universally predict cuspy density
profiles \citep{NFW1996,NFW1997}.

Some recent simulations
\citep[e.g.][]{Governato2012,Pontzen2012,Teyssier2013,DiCintio2014,Brooks2014,Chan2015,Onorbe2015,TrujilloGomez2015,Fitts2017,Maccio2017}
have shown that cores in the central parts of CDM halos can form as a
result of energetic baryon effects, specifically the repeated
injection of supernova energy (following violent episodes of star
formation) into the surrounding gas, the resulting outflows of which
cause DM particle orbits near the halo centre to move out leading to a
new equilibrium system with a central core.

By contrast, the haloes of dwarf galaxies in the \Apostle{}
\citep{Fattahi2016,Sawala2016} and \Auriga{} \citep{Grand2017}
simulations have central cusps, not cores. To investigate the
differences with the simulations that do produce cores, we selected
isolated dwarfs in \Apostle{} and \Auriga{} spanning the mass range
$10^9 < M_{{\rm DM}}/{\rm M}_\odot < 5 \times 10^{10}$. The \Apostle{}
project simulates the formation of the Local Group and its immediate
environment, while the \Auriga{} project consists of re-simulations of
isolated Milky Way-like galaxies. The two sets of simulations differ
in their numerical setups: \Apostle{} was run with a modified version
of the TreeSPH code, \PGadget{}, while \Auriga{} was run with the
moving mesh code, \Arepo{}. Very similar galaxy formation models to
those in \Apostle{} and \Auriga{} have been employed in the larger
scale, cosmological simulations of the \Eagle{} \citep{Schaye2015} and
\Illustris{} \citep{Vogelsberger2014} projects, respectively. These
show that these galaxy formation models lead to {\em galaxy
  populations} which resemble real galaxy populations in many
important properties as a function of time.

Our main conclusions from the current study are:
\begin{enumerate}

\item The size-mass relation of dwarf galaxies in \Apostle{} and
  \Auriga{} exhibits a similar trend to the data for dwarfs in the
  Local Group, albeit with a tighter scatter than what is observed
  (Fig.~\ref{fig:galaxySizes}). For all simulated galaxies with
  stellar mass $M_\star > 10^7 {\rm M}_\odot$, the stellar half-mass
  radius, $r_{1/2}^{\star} > 600$ pc; this is nearly two times larger
  than the nominal resolution limit with which we can reliably measure
  DM profiles from our simulations. Any length scale imposed by the
  formation of these galaxies in the DM density profile would have
  been adequately resolved in both \Apostle{} and \Auriga{}.

\item Irrespective of the amount of stellar mass formed within a dwarf
  galaxy halo, neither \Apostle{} nor \Auriga{} show any evidence of
  core formation. In fact, as shown in Figs.~\ref{fig:flattestAPOSTLE}
  and~\ref{fig:flattestAuriga}, the shallowest inner slope attained by
  the DM density profile of dwarfs in either simulation is $\approx
  -0.8$, far from the slope of 0 corresponding to a constant density
  core.

\item We find no evidence of any correlation between the evolution of
  the inner slope of the DM density profile and the star formation
  rate (SFR) in the \Apostle{} or \Auriga{} dwarf galaxies
  (Fig.~\ref{fig:SFRgamma}); in fact, the evolution of the inner slope
  is consistent with the natural evolution of the inner slope of the
  corresponding haloes in DM-only simulations.

\item Our simulated dwarfs also show no correlation between the
  efficiency of star formation, as measured by $M_\star/M_{{\rm Tot}}$
  (where $M_{{\rm Tot}}$ is the total mass including DM, gas and
  stars), and the change of the inner slope of the DM density profile
  in the hydrodynamics simulations compared to the DM-only cases
  (Fig.~\ref{fig:deltaGammaAPOSTLE}). While the scatter in this
  relation is large, the overall trend is consistent with zero.

\item The star formation histories (SFHs) of a selection of dwarf
  galaxies extracted from \Auriga{} and \Apostle{} (in particular) are
  bursty (Fig.~\ref{fig:SFR}) even when smoothed over 100 and 200 Myr
  intervals (timescales comparable to the typical dynamical time for
  $10^{10} {\rm M}_\odot$ dwarfs at a radius of 1 kpc). The average
  SFRs for these dwarfs can also be quite high, as large as $\sim 3
  \times 10^{-2} {\rm M}_\odot\,{\rm yr}^{-1}$ in some cases.

\item While the SFHs of dwarfs in \Apostle{} are quite bursty and
  those in \Auriga{} less so, dwarfs in both sets of simulations show
  a similar diversity in SFHs when compared to the data for the real
  Local Group dwarfs (Fig.~\ref{fig:sfhAPOSTLE}). In both sets of
  simulations we find examples of dwarfs that range from early to late
  forming, and several that show sustained growth of stellar mass
  throughout their lifetime.

\item The fact that density cores are not generated in these
  simulations, despite the prevalence of bursty SFHs and the
  availability, in principle, of enough energy from supernovae
  feedback, demonstrates that these two conditions are, by themselves,
  {\it insufficient} for core formation.

\end{enumerate}

One possible explanation for the absence of cores is that our
simulations adopt a relatively low gas density threshold for
converting gas into stars which prevents the gas from becoming
gravitationally dominant on kiloparsec scales
(Fig.~\ref{fig:gasDMratio}). However, given the subgrid models
employed in the simulations, this threshold is required to achieve a
good match to the broad population of galaxies. Recent work by
\cite{Read2018} suggests a preference for DM cores in dwarfs that are
gas-rich and highly star forming, compared to a propensity for cusps
in gas-poor, inactive dwarfs. These findings perhaps indicate the
importance for large concentrations of gas over some scale for core
formation to be efficient, for example, the massive gaseous clumps
that e.g. \cite{ElZant2001} and \cite{Nipoti2015} argue can scatter DM
particles away from the centre.

If the presence of density cores at the centres of dwarf galaxies is
eventually established conclusively, this will require an
explanation. One possibility is that the dark matter is more complex
than simple CDM. Another possibility is that the sort of baryon
effects that we have discussed in this paper do, indeed, operate in
nature. It remains to be seen, however, whether a subgrid model can be
constructed which leads to the formation of cores in dwarf galaxies
while preserving the remarkable successes of the \Eagle{} and
\Illustris{} subgrid models in matching properties of the galaxy
population across cosmic time.

\section*{Acknowledgements}
We are grateful to the anonymous referee for providing comments that
have improved this paper. We thank Martin Sparre for suggesting the
check on star formation rate fluctuations in our simulations, and for
other useful discussions. We are also grateful to Matthieu Schaller
and Richard Bower for their comments on an early iteration of this
paper. SB is supported by Harvard University through the ITC
Fellowship, and previously by the STFC through grant ST/K501979/1. RG
acknowledges support by the DFG Research Centre SFB-881 `The Milky Way
System' through project A1. CMS received support from the European
Research Council under ERCStG grant EXAGAL-308037 and the Klaus
Tschira Foundation. CSF acknowledges support from the European
Research Council (ERC) through Advanced Investigator Grant DMIDAS (GA
786910). AF is supported by a European Union COFUND/Durham Junior
Research fellowship (under EU grant agreement no. 609412). JFN
acknowledges the hospitality of the Aspen Center for Physics, which is
supported by National Science Foundation grant PHY-1607611 This work
used the DiRAC Data Centric system at Durham University, operated by
the Institute for Computational Cosmology on behalf of the STFC DiRAC
HPC Facility (\url{www.dirac.ac.uk}). This equipment was funded by BIS
National E-infrastructure capital grant ST/K00042X/1, STFC capital
grant ST/H008519/1, and STFC DiRAC Operations grant ST/K003267/1 and
Durham University. DiRAC is part of the National
E-Infrastructure. This research was carried out with the support of
the HPC Infrastructure for Grand Challenges of Science and Engineering
Project, co-financed by the European Regional Development Fund under
the Innovative Economy Operational Programme. This research was
supported in part by the National Science Foundation under Grant
No. NSF PHY17-48958. The data analysed in this paper can be made
available upon request to the author.

\bibliographystyle{mnras}
\bibliography{profiles.bib}{}

\label{lastpage}

\end{document}